\pdfoutput=1
\documentclass[12pt]{article}
\usepackage{jheppub}

\usepackage[utf8]{inputenc}

\usepackage{amsmath}
\usepackage{amssymb}
\usepackage{graphicx,color}
\usepackage{amsthm}
\usepackage{mathrsfs}

\usepackage{tikz}
\usetikzlibrary{matrix,arrows}

\usepackage{ifpdf}

\usepackage{subfigure}

\usepackage[boxsize=0.5em,aligntableaux=center]{ytableau}

\usepackage{tikz}
\usetikzlibrary{shapes,shadows,calc}
\usepgflibrary{arrows}
\usepackage{rotating}
\usepackage{wrapfig}

\tikzset{
  sshadow/.style={opacity=.25, shadow xshift=0.05, shadow yshift=-0.06},
}

\def\kbbox[#1,#2,#3,#4,#5]#6{
        \draw[dashed] node[draw,color=gray!50,minimum
        height=#1,minimum width=#2] (#4) at #5 {}; 
        \node[anchor=#3,inner sep=2pt] at (#4.#3)  {#6};
}
\def\kbboxred[#1,#2,#3,#4,#5]#6{
        \draw[] node[draw,color=red,minimum
        height=#1,minimum width=#2] (#4) at #5 {}; 
        \node[anchor=#3,inner sep=2pt] at (#4.#3)  {#6};
}

\newcommand{\cB}{\mathcal{B}}
\newcommand{\cC}{\mathcal{C}}

\newcommand{\cK}{\mathcal{K}}

\newcommand{\cM}{\mathcal{M}}
\newcommand{\cN}{\mathcal{N}}

\newcommand{\cV}{\mathcal{V}}

\def\a{{\alpha}}
\def\b{{\beta}}

\def\lam{{\lambda}}

\def\g{{\gamma}}
\def\k{{\kappa}}

\def\p{{\partial}}
\def\l{{\langle}}
\def\r{{\rangle}}

\def\bs#1\es{\begin{split}#1\end{split}}
\def\ba#1\ea{\begin{align}#1\end{align}}
\def\baed#1\eaed{\begin{aligned}#1\end{aligned}}
\def\bged#1\eged{\begin{gathered}#1\end{gathered}}

\def\nn{\nonumber}

\def\a{\alpha}
\def\b{\beta}

\def\d{\delta}
\def\D{\Delta}

\def\F{\Phi}
\def\g{\gamma}

\def\h{\eta}
\def\k{\kappa}
\def\l{\lambda}
\def\L{\Lambda}
\def\m{\mu}
\def\n{\nu}
\def\o{\omega}

\def\Q{\Theta}
\def\r{\rho}
\def\s{\sigma}
\def\S{\Sigma}

\def\x{\xi}

\def\bls{\bigg [}
\def\brs{\bigg ]}

\def\pa{\partial}

\def\fr{\frac}

\def\we{\wedge}

\def\inv{{\check{k}}}
\def\Pperp{{\Pi_{\perp}{}}}

\newcommand{\til}[1]{ {\tilde{#1}} }

\let\foo\bar 
\renewcommand{\bar}[1]{ {\foo{  #1} }{} }

\newlength{\dhatheight}

\def\ihat{{\hat{\textrm{\emph{\i}}}}}
\def\jhat{{\hat{\textrm{\emph{\j}}}}}

\def\Label#1{\label{#1}%
  \smash{\hbox to0pt{\raise1ex\hbox{\tiny[#1]}\hss}}}
\def\noLabels{\let\Label=\label}
\def\nobbibitem{\let\bbibitem=\bibitem}
 \def\noBibitem{\let\Bibitem=\bibitem}

\newcommand{\be}{\begin{equation}}
\newcommand{\ee}{\end{equation}}
\newcommand{\beq}{\begin{equation}}
\newcommand{\eeq}{\end{equation}}
\newcommand{\bea}{\begin{eqnarray}}
\newcommand{\eea}{\end{eqnarray}}
\newcommand{\R}{\text{Re}}
\newcommand{\I}{\text{Im}}

\newcommand{\w}{\wedge}
\newcommand\varpm{\mathbin{\vcenter{\hbox{%
  \oalign{\hfil$\scriptstyle+$\hfil\cr
          \noalign{\kern-.3ex}
          $\scriptscriptstyle({-})$\cr}%
}}}}
\newcommand\varmp{\mathbin{\vcenter{\hbox{%
  \oalign{$\scriptstyle({+})$\cr
          \noalign{\kern-.3ex}
          \hfil$\scriptscriptstyle-$\hfil\cr}%
}}}}

\title{\centering \Large Non-Abelian discrete gauge symmetries in F-theory}

\author[]{Thomas W.~Grimm,}
\author[]{Tom G.~Pugh,}
\author[]{and Diego Regalado}

\affiliation[]{Max Planck Institute for Physics,\\
F\"ohringer Ring 6, 80805 Munich, Germany}

\emailAdd{grimm@mpp.mpg.de}
\emailAdd{pught@mpp.mpg.de}
\emailAdd{regalado@mpp.mpg.de}

\abstract{The presence of non-Abelian discrete gauge symmetries in four-dimensional F-theory compactifications
               is investigated. Such symmetries are shown to arise from seven-brane configurations in genuine F-theory  
               settings without a weak string coupling description.
               Gauge fields on mutually non-local seven-branes are argued to gauge both R-R and NS-NS 
               two-form bulk axions. The gauging is completed into a generalisation of the Heisenberg group
               with either additional seven-brane gauge fields or R-R bulk gauge fields. The former case
               relies on having seven-brane fluxes, while the latter case  
               requires torsion cohomology and is analysed in detail through the M-theory dual. Remarkably,
               the M-theory reduction yields an Abelian theory that becomes non-Abelian
               when translated into the correct duality frame to perform the F-theory limit. 
               The reduction shows that the gauge coupling function depends on the 
               gauged scalars and transforms non-trivially as required for the 
               groups encountered. 
                This field dependence agrees
               with the expectations for the kinetic mixing of seven-branes and is unchanged 
               if the gaugings are absent.  
               }

\begin{document}
\setlength{\parskip}{5pt}

\makeatletter
\renewcommand{\@fpheader}{%\old@fpheader
\hfill MPP-2015-41}
\makeatother

\maketitle
\newpage
\section{Introduction}

In recent years phenomenological aspects of F-theory compactifications \cite{Vafa:1996xn} have 
been considered intensively. While a complete understanding of 
the effective actions arising in such compactifications is still lacking there has been 
major progress  investigating core aspects of the  theories that arise. Much of 
these efforts have focused on uncovering the geometric manifestation 
of symmetries of the effective theories in F-theory. For example, a detailed 
picture of the local continuous non-Abelian and Abelian gauge symmetries 
has started to emerge. A state-of-the-art discussion on non-Abelian symmetries in F-theory can be 
found in \cite{Hayashi:2014kca,Braun:2014kla,Esole:2014bka,Esole:2014hya}, while recent results on Abelian gauge symmetries are found in \cite{Grimm:2010ez,Morrison:2012ei,Braun:2013yti,Grimm:2013oga,Kuntzler:2014ila,Borchmann:2013hta,Cvetic:2013nia,Cvetic:2013uta,Borchmann:2013jwa,Cvetic:2013qsa,Braun:2013nqa,Klevers:2014bqa,Braun:2014qka,Lawrie:2014uya,Lawrie:2015hia,Grimm:2015zea}. The investigation of discrete symmetries in F-theory has only recently 
attracted more attention \cite{Braun:2014oya,Morrison:2014era,Anderson:2014yva,Klevers:2014bqa,Garcia-Etxebarria:2014qua,Mayrhofer:2014haa, Mayrhofer:2014laa,Cvetic:2015moa}. The class of discrete symmetries are thereby 
realised as low energy remnants of Abelian gauge symmetries that are massive 
even in the absence of any flux background. This makes these symmetries necessarily
Abelian. In this work we aim to generalize the results of \cite{Anderson:2014yva,Garcia-Etxebarria:2014qua,Mayrhofer:2014laa} and 
discuss the appearance of non-Abelian discrete symmetries in F-theory.

In F-theory various aspects of the physics of intersecting seven-branes are captured by higher-dimensional two-torus fibered geometries.  
It turns out, however, that extracting the low energy implications of a given 
geometry is a challenging task. This can be traced back to the fact, that there is no known 
twelve-dimensional formulation of F-theory and the theory has to be studied either 
by a generalised Type IIB string perspective or by performing a duality to M-theory.
Approaching F-theory directly from the
Type IIB perspective seems to avoid the use of any dualities. 
However, as of now the correct global treatment of intersecting 
seven-branes is poorly understood and can be very involved, 
even in simple higher-dimensional compactifications.
This is particularly apparent when dealing with two or more 
seven-branes that are mutually non-local, i.e.~they cannot 
be rotated to D7-branes by the $Sl(2,\mathbb{Z})$ symmetry at the same time.\footnote{A recent example of this fact is analysed in \cite{Douglas:2014ywa}.}
As we will find in this work, this is precisely the kind of brane configurations 
that can realise certain non-Abelian discrete symmetries. Such 
situations are better understood using the duality to M-theory.
In this case, however, one also has to face a major complication. 
Since generating global discrete symmetries in 
a theory of quantum gravity requires that these 
are obtained from broken local gauge symmetries (see, e.g.~\cite{Banks:2010zn}), one typically 
has to have a proper treatment of massive states in the effective action. 
This can be involved when using the M-theory to F-theory limit, 
since one needs to disentangle whether a mass of a state is actually present 
in F-theory or is a remnant of the fact that the F-theory limit has not been performed. 

A key example of the complications which arise when dealing with massive modes in F-theory 
is given by so-called `geometrically massive' U(1) gauge symmetries discussed in \cite{Grimm:2010ez,Grimm:2011tb,Grimm:2013fua,Braun:2014nva}. Such massive U(1)s are 
familiar from Type IIB orientifold compactifications, where they arise from specific 
configurations of D7-branes and their orientifold images \cite{Jockers:2004yj}. 
The brane U(1)s are massive even in 
the absence of brane fluxes with a mass proportional to the string coupling. 
Leaving the Type IIB weak string coupling limit requires one to realise such massive 
U(1)s via a torus-fibered geometry used in M-theory. Geometrically massive 
U(1)s are then believed to arise from the expansion into non-harmonic forms. 
These forms might be described by non-trivial torsion in cohomology \cite{Camara:2011jg,Grimm:2011tb,BerasaluceGonzalez:2012vb,Mayrhofer:2014laa}, 
but eventually require the M-theory geometries considered to be extended to include non-K\"ahler spaces \cite{Grimm:2010ez,Grimm:2011tb,Grimm:2013fua,Braun:2014nva}. Remarkably, this allows these U(1)s to mix also with the 
Kaluza-Klein vector used in connecting F-theory and M-theory. It was argued in \cite{Anderson:2014yva,Garcia-Etxebarria:2014qua,Mayrhofer:2014laa}
that this is the proper interpretation of the physics induced 
on certain torus-fibered geometries with multi-section. 
Such massive U(1)s were argued to lead to interesting 
discrete Abelian symmetries restricting, for example, the Yukawa couplings of
the effective theories \cite{Garcia-Etxebarria:2014qua,Mayrhofer:2014haa,Cvetic:2015txa}.  

Given the success of identifying at least certain 
discrete Abelian symmetries in the F-theory geometry one might hope to be able 
to straightforwardly generalize the setting to the non-Abelian case. However, this 
leads immediately to some obstacles. Firstly, the study of 
geometries with multi-section seems to suggest that only Abelian symmetries 
naturally appear in such settings.\footnote{Indeed, it has been proposed 
that the Tate-Shafarevich group determines the discrete symmetries arising in F-theory \cite{Braun:2014oya}. 
This group, however, is always Abelian and therefore obscures any generalisation.}
Secondly, as we will see in more detail below, including non-closed forms 
in the reduction of M-theory accessed via eleven-dimensional supergravity
seemingly only yields Abelian gaugings. How can a non-Abelian discrete 
symmetry ever arise? This appears particularly puzzling, since 
we know from the analysis of the Type IIB supergravity actions that 
non-Abelian discrete symmetries actually do occur in reductions 
with non-closed forms representing torsion cohomology \cite{Gukov:1998kn,BerasaluceGonzalez:2012vb}. 
These arise from gauging a subgroup of the isometry group of the moduli space and 
are known to span a generalisation of the Heisenberg group.
%and therefore should be embeddable into F-theory studied via M-theory. 
In this work we resolve these puzzles by explaining that they can be traced 
back to the fact that the M-theory reduction is performed in an inconvenient
duality frame working with the M-theory three-form $C_3$ only. While 
the gaugings in the M-theory reduction with $C_3$ 
appear to be Abelian they actually dualize into 
non-Abelian gaugings of the Heisenberg algebra in the duality frame required 
to perform the F-theory limit.

Our findings admit an interesting Type IIB interpretation using the 
geometric St\"uckelberg mechanism. Recall that on a D7-brane this 
mechanism only allows one to gauge the R-R two-form axion with the brane 
U(1). This implies that the $Sl(2,\mathbb{Z})$-images generally allow for a gauging of the NS-NS two-form axions as well. If one now includes two seven-branes 
that are mutually non-local we will show that this can imply that a third vector  
has to complete the gauging into a non-Abelian group. 
This additional vector can arise either from the R-R bulk sector, in which case 
the gaugings are purely geometrical, or from another seven-brane, in which 
case brane fluxes are required. We argue that the former possibility admits a 
direct interpretation in the M-theory fourfold geometry. The non-Abelian completion 
of the gauging turns out to be a consistency condition on the compact fourfold when considering 
non-closed forms. It admits a natural mathematical interpretation in terms of 
 torsion cohomology for Calabi-Yau fourfolds.  

The non-Abelian gauge groups that we find are shown to be generalisations 
of the Heisenberg group. The fact that these groups are neither semisimple 
nor compact has important consequences on the form of the effective action.
It implies that the gauge coupling function of this group cannot be 
simply constant, since there exists no positive definite Killing form for these groups. 
It instead has to be a non-trivial holomorphic function of the complex scalars transforming 
under the gauge group. Interestingly the gauge coupling function is independent 
of the constants determining the gauged subalgebra of the isometry group. 
This implies that they are present for any gauging and we argue that they 
can be used to determine the allowed non-Abelian gauge algebra. 

The paper is organised as follows. In section \ref{sec:review} we review how  
non-Abelian discrete symmetries can arise as gaugings of isometries in 
four dimensions. The Type IIB string theory embedding of a special 
type of gaugings is discussed in section \ref{IIBorientifold}. We show that when using seven-brane
gauge fields to yield such gaugings the introduction of mutually non-local 
seven-branes is crucial. This suggests that a proper treatment should  
invoke an F-theory geometry and the duality to M-theory. The F-theory setting and the allowed gaugings are discussed in section \ref{nonA_inFtheory}, while the 
explicit M-theory reduction is then performed in section \ref{MFduality}. We 
show that the Abelian gaugings dualize to non-Abelian 
gaugings upon changing to the duality frame that allows  the 
F-theory limit to be performed. Details on the computations are supplemented in appendix \ref{DualAction}.

\section{Non-Abelian discrete symmetries in four dimensions}\label{sec:review}

In this section we briefly review the realisation of discrete gauge symmetries in field theory \cite{Banks:2010zn,BerasaluceGonzalez:2011wy,BerasaluceGonzalez:2012vb}. We also include a discussion of supersymmetry and comment 
on the structure of non-minimal gauge-kinetic terms for non-Abelian groups that are 
neither semisimple nor compact. 

\subsection{Non-Abelian discrete symmetries}

By discrete gauge symmetry we simply mean a discrete remnant of a spontaneously broken gauge symmetry. Let us consider the simplest Abelian example to illustrate this, namely the St\"uckelberg Lagrangian for a vector $A$ and a scalar $\phi$ of periodicity $2\pi$,
\be\label{stuck}
\mathcal L= -\frac{1}{2g^2}dA\wedge *dA-\mu^2(d\phi-kA)\wedge *(d\phi-kA) \, , 
\ee
where $g$ is the YM coupling constant, $\mu$ is a mass scale and $k\in \mathbb Z$. This Lagrangian is invariant under the local transformations
\ba
\delta A&=d\lambda, & 
\delta\phi&= k\lambda
\ea
and we find that the space of physically distinct vacua is given by $A=0$ and $\phi=\phi_0$ with $\phi_0$ a constant in $[0,2\pi)$. Then, we immediately see that this system breaks the underlying $U(1)$ symmetry since under a constant gauge transformation we find that the vacuum is not invariant. Indeed, if we consider the vacuum defined by $|\phi_0\rangle$, then after such gauge transformation we arrive at $|\phi_0+k\lam\rangle$, which is in general different from $|\phi_0\rangle$. However, due to the presence of the integer $k$ and the periodicity of $\phi$, we may still find non-trivial gauge transformations that preserve the vacuum, namely $\lam=\frac{2\pi}{k}$, which form a $\mathbb Z_k$ subgroup of $U(1)$ parameterised by $e^{i\lambda}$. 
Let us stress that the fluctuations of the vector $A$ around these vacua is 
massive with mass $k^2 \mu^2$. This implies that an effective theory 
arising from string theory has to include massive modes. 

As shown in \cite{BerasaluceGonzalez:2012vb}, in order to generalise this to non-Abelian discrete symmetries, it proves useful to think of (\ref{stuck}) as the gauging of a scalar manifold with a $U(1)$ isometry with charge $k$. In that case we start with a scalar manifold $S^1$, whose isometry group is generated by $t=\p_\phi$. Furthermore, the particular gauging we consider is related to picking a Killing vector with the following normalisation
\be\label{killab}
X=k \pa_\phi.
\ee
Then, the orbit associated to $X$ is a map $Q: S^1\times U(1)\rightarrow  S^1$ that takes a point $\phi_0\in S^1$ and the element $e^{i\lambda}\in U(1)$ to give $Q(\phi_0;\lambda)=\phi_0+k\lambda$. Then, we see that for a given vacuum $\phi_0$, the subgroup that is not broken corresponds to the solutions to $Q(\phi_0;\lambda)=\phi_0$ which again leads to $\mathbb Z_k$. Notice that the discrete symmetry is encoded in the relative normalisation of the Killing vector (\ref{killab}) with respect to the gauge algebra generator.

Next we discuss the gauging of non-Abelian isometries with 
the appropriate charges. 
Consider a sigma model with a $d$-dimensional manifold $\mathcal M$ endowed with a Riemannian metric $g$ and coordinates $\phi^a$,
\be
\mathcal L_0=- g_{ab}d\phi^a\wedge *d\phi^b,
\ee
and let $t_{\hat A}$ be generators of the group of isometries Iso$(\mathcal M)$ which satisfy 
\beq \label{non-Ab_structure}
[t_{\hat A},t_{\hat B}]=f_{\hat A\hat B}{}^{\hat C} t_{\hat C}\ ,
\eeq 
where $f_{\hat A\hat B}{}^{\hat C}$ are the structure constants. 
A particular gauging is specified by picking a set of Killing vectors 
\ba
X_A=k_A^{\hat A}t_{\hat A}
\ea
where $k_A^{\hat A}$ are constants and the vectors generate the gauge algebra
\ba
[X_A,X_B]=f_{AB}{}^CX_C.
\ea
The gauging is implemented by considering the following Lagrangian
\be\label{gaugednon}
\mathcal L=-\frac{1}{2}f^1_{AB}F^A\wedge *F^B-\frac{1}{2}f^2_{AB}F^A\wedge F^B- g_{ab}D\phi^a\wedge *D\phi^b \ , 
\ee
where we included gauge bosons $A^B$ with field strengths $F^B=dA^B+f_{AC}{}^BA^A\wedge A^C$. The functions $f^1_{AB}$ and $f^2_{AB}$ are in general dependent on 
the scalars $\phi^a$. $f^1_{AB}$ determines the gauge couplings and has to be positive definite.
We stress that $f^i_{AB}, \, i=1,2$ in general have to transform non-trivially  
under the gauge group in order to ensure gauge invariance of the Lagrangian, i.e.~one 
has to have 
\beq  \label{f-transform}
    \delta f^i_{AB} = \lambda^C (f_{CA}{}^D f^i_{BD} + f_{CB}{}^D f^i_{AD}) \ ,
\eeq
where $\lambda^D$ are the gauge parameters. 
In particular, for groups that are noncompact one cannot use 
the Killing form and therefore $f^1_{AB}$ and $f^2_{AB}$ have to be non-trivial 
functions of the fields $\phi^a$. 
Furthermore, we defined the covariant derivatives
\be
D\phi^a=d\phi^a-A^BX_B^a.
\ee
Now we may proceed formally in analogy to the Abelian case. The space of inequivalent vacua of the gauged theory (\ref{gaugednon}) is $A^B=0$ and constant $\phi^a\in \mathcal M$. Then, under a constant gauge transformation along $\lam^AX_{ A}$ we find that the vacuum $\phi_0^a$ goes to $Q(\phi_0^a;\lam^A)$ which, if different from $\phi_0^a$, signals a spontaneous breaking of the generator $\lam^AX_A$. Alternatively, the set of $e^{i\lam^{ A} X_{ A}}$ that satisfy $Q(\phi_0^a;\lam^{ A})=\phi_0^a$ corresponds to a preserved symmetry. Clearly, this construction may lead to a case in which the preserved symmetry is a non-Abelian discrete subgroup of Iso$(\mathcal M)$. In the following section we consider a particular example.

\subsection{Supersymmetric non-Abelian gaugings and non-minimal kinetic terms} \label{non-min-kinetic-terms}

Up to now we did not discuss the supersymmetric version of the above setting. In order to do that one has to first realise 
that four-dimensional $\cN=1$ supersymmetry implies that $\cM$ is 
a K\"ahler manifold. We denote the complex coordinates by $\Phi^I$.
The couplings $f^1_{AB}$ and $f^2_{AB}$ have to combine into a function $f_{AB} = f^1_{AB} + i f^2_{AB}$ that is holomorphic in the $\Phi^I$. The bosonic part of a supersymmetric 
Lagrangian will then include the terms 
\beq \label{N=1lagrangian_gen}
 \mathcal L=-\frac{1}{2}\R f_{AB}F^A\wedge *F^B-\frac{1}{2}\I f_{AB}F^A\wedge F^B-
  K_{I \bar J}D\Phi^I\wedge *D\bar \Phi^J\ -V*1,
\eeq
where $K_{I\bar J} = \partial_{\Phi^I} \partial_{\bar \Phi^J} K$ is locally the derivative of a K\"ahler potential $K$.
Crucially the isometries that can now be gauged have to be holomorphic such that 
\ba
\d A^A & =  d \L^A + f_{B C}{}^A A^B \L^C & 
\d \F^I &= \L^A X_A{}^I ( \F )  &
\ea
which induces the transformation  \eqref{f-transform}.

Let us stress that in general one has to impose additional conditions on gaugings allowed 
by $\cN=1$ supersymmetry. Consider, for example, the simple K\"ahler potential $K = \frac{1}{2}(\phi + \bar \phi)^2$,
which yields a constant K\"ahler metric and a Lagrangian that admits the shift symmetry $\phi \rightarrow \phi +\lambda_1 + i \lambda_2$ with real constants $\lambda_i$. Supersymmetry implies that the two shift symmetries labelled by $\lambda_i$ cannot be gauged by different $U(1)$ vectors $A^i$, since the D-term scalar potential would not be gauge-invariant. In our examples the situation will be even more subtle. Since the gaugings also have to be compatible 
with the holomorphicity of the gauge coupling function even if there exists a gauge-invariant scalar potential. 

If the isometry group Iso$(\mathcal M)$ that we want to gauge is semisimple and compact,  we may take $f_{AB}$ 
to be a holomorphic function of the ungauged scalars proportional to the Killing form. In this case $f_{AB}$ satisfies the 
constraints imposed by supersymmetry. This implies that one can also add the kinetic terms for $A^A$
to the Lagrangian without having any gauged scalars. 
However, the isometry group need not be compact nor semisimple in which case we might be forced to include non-minimal kinetic terms for the vectors. In such cases, holomorphicity of the gauge kinetic function imposes non-trivial constraints \cite{Hull:1985pq}.

%In order to make this more precise, one first notes that, at least locally, $\cM = \cM_0 \times \text{Iso} \cM/H$, where $H$ is the centralizer of Iso$\cM$
%and $\cM_0$ is not acted upon by the isomenties. This gives a local embedding 
%of isometry group into $\cM$. Now one can integrate 
%\eqref{f-transform} and consider $f_{AB}$ on the above local 
%split. One thus finds it to take the form
%\beq \label{gauge-nonsemi}
    % f_{AB}(\Phi)  = f_{CD}^0(\Phi) D^{-1 C}_A(\Phi)  D^{-1 D}_B(\Phi) \ ,
%\eeq
%where $D^{\ C}_A$ is the adjoint representation matrix of Iso$\cM$ embedded 
%into $\cM$. The holomorphic function $f_{CD}^0$ is required to be invariant 
%under $H$. 
%If the killing vectors $X_A{}^I \pa_{ \F^I}$  become linearly dependent over the complex numbers at some point in the moduli space then the integration of \eqref{f-transform} allows $D^{-1 C}_A$ to become non-holomorphic and supersymmetry may be broken. For a given $\cM$ the determination of the gauge couplings $f_{AB}$ 
%compatible with all the constraints can be involved and we refer the reader to \cite{xxx}
%for more details. 

%We will show in the next sections that the 
%relevant examples to obtain non-Abelian discrete symmetries  
%have a non-semisimple isomety group and couplings are indeed of the form \eqref{gauge-nonsemi}.
%It turns out, however, that the group $H$  is trivial in our examples and $f^0_{AB}$
%is a general holomorphic function of the ungauged fields. 

Let us close this section with recalling yet another important issue related to the 
gauge-transformation of the gauge coupling function. If the four-dimensional $\cN=1$
theory contains chiral fermions charged under an Abelian gauge symmetry, 
it might be necessary to employ a Green-Schwarz mechanism to chancel the one-loop 
anomaly induced by these fields \cite{Harvey:2005it,Bilal:2008qx}. The classical terms $\frac{1}{2}\I f_{AB}F^A\wedge F^B$ 
are then allowed to be non-gauge invariant and fixed to induce tree-level diagrams 
that cancel the one-loop anomalous diagrams of the chiral fermions. In consistent string theory 
compactifications this mechanism is automatically implemented in the situations that 
require such a cancellation.

\section{Non-Abelian discrete symmetries in Type IIB orientifolds}\label{IIBorientifold}

In this section we study the possibility of obtaining non-Abelian discrete symmetries 
by gauging R-R and NS-NS scalars in Type IIB orientifolds with O7-planes. 
We first examine the symmetries of the orientifold moduli space in 
subsection \ref{Heissenberg_ori}. The Heisenberg isometry group that appears 
is a special version of the symmetry groups later encountered in the complete F-theory setting. 
We then turn to the discussion of the gauging of this non-Abelian group in subsection \ref{susygaugings_ori}
by performing a reduction with non-harmonic forms. It turns out that there is a tension between 
performing a supersymmetric orientifold quotient and the gauging of a non-Abelian group.

\subsection{Heisenberg isometries in Type IIB orientifold compactifications} \label{Heissenberg_ori}

To begin with let us consider Calabi-Yau orientifold compactifications of Type IIB with O7-planes.
The effective action for the bulk fields in such compactifications
contains the following terms \cite{Grimm:2004uq}
\bea\label{schem_text}
\mathcal L&=&-G_{\a\b}dv^\a\w* dv^\b-\frac{1}{4 \cV^2}d \cV\w*d\cV
+ \frac{3iv^\a}{(\tau-\bar\tau)}\mathcal K_{\a ab}(dc^a-\tau db^a)\w*(dc^b-\bar\tau db^b) \nn \\
&&-\frac{G^{\a\b}}{16 \cV^2}\Big(d\rho_\a+\frac{1}{2}\mathcal K_{\a ab}(b^adc^b-c^adb^b)\Big)\w*\Big(d\rho_\b+\frac{1}{2}\mathcal K_{\b cd}(b^cdc^d-c^cdb^d)\Big)\ .
\eea
In this expression $\tau = C_0 + i e^{-\phi}$ is the axiodilaton, 
$b^a,c^a,\, a=1,\ldots h^{1,1}_-$ arise from the reduction of $B_2$ and $C_2$ on harmonic orientifold-odd 
two-forms, and $\rho_\alpha,\, \alpha=1,\ldots,h^{1,1}_+$ comes from the reduction of $C_4$ on orientifold-even harmonic four-forms.
The real scalars $v^\alpha$ are the deformations of the K\"ahler form of 
the underlying Calabi-Yau geometry. The intersection numbers of the Calabi-Yau manifold 
are given by 
\ba
\cK_{\alpha \beta \gamma} & = \int_{Y_3} \o_\a \we \o_\b \we \o_\g \, , &
\cK_{\alpha ab} & = \int_{Y_3} \o_\a \we \o_a \we \o_b \, . 
\ea
The first of these is related to the definition of the volume $\cV= \frac{1}{6} \cK_{\alpha \beta \gamma} v^\alpha v^\beta v^\gamma$ and the metric $G_{\alpha \beta}$. The Lagrangian defines a K\"ahler metric when written 
in the form \eqref{N=1lagrangian_gen} with a K\"ahler potential $K = - 2 \log \cV$
and complex coordinates 
\beq
   G^a = c^a - \tau b^a \ , \qquad T_\alpha  = \rho_\alpha + \frac{1}{2(\tau -\bar \tau)} \cK_{\alpha ab} G^a (G-\bar G)^b - \frac{1}{2} i \cK_{\alpha \beta \gamma} v^\beta v^\gamma\ .
\eeq
Clearly, there will be additional moduli corresponding to the complex structure 
deformations and brane fields. These will suppressed in the following, since our current 
focus is on the identification of candidate non-Abelian symmetries in this sector of the 
theory. As we will see later, similar structures appear in the seven-brane sector. 

One now readily checks that this K\"ahler metric has 
the following $2h^{(1,1)}_-+h^{(1,1)}_+$ holomorphic isometries
\ba
\label{GTisom}
\delta G^a &=\,\,\lam_1^a-\tau\lam_2^a\ , &
\delta T_\a &=\,\,\lam_\a-\mathcal K_{\a ab}G^b\lam_2^a\ .
\ea
where $\lam_1^a,\lam_2^a,\lam_\a$ are the real scalar gauge parameters. 
Using the transformations \eqref{GTisom} one determines the holomorphic 
Killing vectors to be 
\ba
    t_{(1,a)} &=  \partial_{G^a}  \ , & 
     t_{(2,a)} &= - \tau \partial_{G^a} - \cK_{\alpha a b} G^b \partial_{T_\alpha} \ , &  
    t^{\alpha} &= \partial_{T_\alpha}\ .
\ea 
Upon exponentiation these vectors yield the Lie group of isometries of $\mathcal M$, which we denote by ${\rm Iso(\mathcal M)}$. The explicit algebra reads, 
\beq \label{Heisenberg_complex}
  [ t_{(1,a)},   t_{(2,b)}]=-\mathcal K_{\a ab} t^{\a} \ , 
\eeq
with all other commutators vanishing. This algebra is a generalisation of the 
Heisenberg algebra and will be our prime example for the non-Abelian 
structures appearing in our string theory set-ups.
Comparing with \eqref{non-Ab_structure} this implies that the only non-vanishing non-Abelian structure 
constants are $f_{(1,a)(2,b)}{}^{\a}=-\mathcal K_{\a ab}$.
Finally, the fact that $c^a,b^a$ and $\rho_\a$ are periodic with period $2\pi$, imposes the following identifications in the scalar manifold
\ba
\label{ident}
c^a\simeq&\,\, c^a+2\pi\, ,\quad \text{and}\quad \rho_\a\simeq \rho_\a+\pi\mathcal K_{\a ab}b^b\, , \nn \\
b^a\simeq& \,\, b^a+2\pi\, ,\quad \text{and}\quad \rho_\a\simeq \rho_\a-\pi\mathcal K_{\a ab}c^b\, , \nn \\
\rho_\a\simeq&\,\, \rho_\a+2\pi\, .
\ea
These identifications render the field-space spanned by $c^a,b^a$ and $\rho_\alpha$ to be 
compact. 

Let us now address the question of gauging the non-Abelian 
symmetries \eqref{Heisenberg_complex}. This requires the introduction 
of gauge fields that arise from the bulk sector. In section \ref{nonA_inFtheory} we will develop this further by including vectors that arise from the brane sector. 

\subsection{Non-Abelian gaugings from Type IIB orientifolds with torsion} \label{susygaugings_ori}

In this section we briefly review the construction in \cite{BerasaluceGonzalez:2012vb} which shows that the reduction of Type IIB on manifolds $Y_3$ with torsion homology may lead
to an effective theory where the non-Abelian isometries analysed in the last section are gauged. 

In general, cohomology groups with integer coefficients are finitely generated Abelian groups, which means that they are the direct sum of cyclic groups, namely
\be
H^p(M,\mathbb Z)=\underbrace{\mathbb Z\oplus\cdots\oplus\mathbb Z}_{\rm free}\oplus \underbrace{\mathbb Z_{k_1}\oplus\cdots\oplus\mathbb Z_{k_n}}_{\rm torsion}
\ee
which, as indicated above, is the sum of a free part and a non-free (or torsion) part. The former plays a central role in string compactifications since Hodge's theorem
provides an isomorphism between the free part and the space of harmonic forms, which correspond to the internal wave function of massless modes. The torsion part, however,
does not yield massless modes so its role in compactifications is not as straightforward. It was argued in \cite{Camara:2011jg,BerasaluceGonzalez:2012vb} that including torsion forms in string reductions 
naturally yields discrete gauge symmetries. Also, one can obtain the correct spectrum of charged states under such discrete symmetry by
wrapping different branes in the torsion homology cycles, in agreement with the expectations for a theory of quantum gravity \cite{Banks:2010zn}.

Let us now illustrate the reduction on torsion cohomology in a simple example before moving to a more general case. We will consider the  reduction of
a theory with a two form potential $B_2$ to four dimensions on a space $M$ with torsion cohomology ${\rm Tor }H^2(M,\mathbb Z)=\mathbb Z_k$. Then we have
a closed two-form $\Lambda_2$ which, in integer cohomology, is not exact but such that $k$ times $\Lambda_2$ is, namely 
\ba\label{torsionform}
k\Lambda_2=d\lambda_1
\ea
for some form $\lambda_1$. Now if we include the torsion
element $\Lambda_2$ in the reduction, then we must also reduce along the non-closed form $\lambda_1$. This follows from the fact that the Laplacian $\Delta$ commutes with the exterior derivative and
from demanding that we include all the modes of a given mass scale. Indeed, if $\Delta\,\Lambda_2=-m^2\Lambda_2$, we should include the form $\lambda_1$ which satisfies that
$\Delta \,\lambda_1=-m^2 \lambda_1$. Thus, we find that
\ba
B_2&=b\,\Lambda_2+A\w\lambda_1, & dB_2&=(db-kA)\w\Lambda_2+dA\w\lambda_1
\ea
where $b$ and $A$ are four-dimensional scalar and vector which appear in the field strength for $B_2$ in the combination $(db-kA)$ which gives a St\"uckelberg coupling.
This then leads to a theory of a massive vector with a $\mathbb Z_k$ discrete gauge symmetry.

Now we are ready to discuss the more involved case of Type IIB orientifolds with torsion. A six-dimensional manifold has only two independent torsion cohomology groups, namely
\ba
%{\rm Tor }H_1(Y_3)\simeq {\rm Tor }H_4(Y_3)\simeq 
{\rm Tor } H^2(Y_3)&\simeq {\rm Tor }H^5(Y_3)\simeq \bigoplus_a \mathbb Z_{k_a},&
%{\rm Tor }H_2(Y_3)\simeq {\rm Tor }H_3(Y_3)\simeq
 {\rm Tor } H^3(Y_3)&\simeq {\rm Tor }H^4(Y_3)\simeq \bigoplus_\a \mathbb Z_{k_\a}
\ea
where the isomorphisms follow from the universal coefficient theorem. Then, in analogy with equation (\ref{torsionform}) we have that\footnote{We did not include the torsion five-forms since we will not need them here.}
\ba 
   d\gamma_i &= k_{i}^a \omega_a\ \, ,&
   d \omega_\alpha &= k_{\alpha \kappa} \beta^{\kappa}\ , &
   d\alpha_\kappa &= k_{ \alpha\kappa} \tilde \omega^\alpha \, , 
   \nn \\
   d \o_a & = 0 \,, & 
   d \til \o^\a & = 0 \, , & 
   d\b_\kappa &= 0 
   \label{domega_ori}
\ea
which are compatible with the conditions 
\ba
\int_{Y_3} \alpha_\kappa \wedge \beta^\lambda &=\delta_{\kappa}^\lambda \, , & 
\int_{Y_3} \omega_\alpha \wedge \tilde \omega^\beta &=\delta_{\alpha}^{\beta}.
\ea
We note that in the pure torsion case the $k_{i}^a$ and $k_{ \alpha\kappa}$ would be invertible. However, by not imposing conditions on the rank we allow harmonic and torsion forms to be considered simultaneously in the following analysis. Also, we assume that the parity under the orientifold action of $\alpha_\kappa,\beta^\kappa$ and $\o_\a,\tilde\o^\a$ is even while the parity of $\gamma_i$ and $\o_a$ is odd. 

In addition to this we will also demand that the basis of forms also satisfies 
\ba\label{complete}
\o_a \we \g_i & = M_{i a}{}^\k \a_\k \, ,&
\o_a \we \o_b & = \cK_{\a a b} \til \o^\a \, ,&
\g_i \we \g_j & = N_{i j}{}^\a \o_{\a} \, . 
\ea
In the first of these identities we have demanded that there is no term proportional to $\b^\k$. This is imposed in order to prevent electric and magnetic degrees of freedom from being simultaneously gauged. The quantities $M_{i a}{}^\k $  and $N_{i j}{}^\a $ appearing in these identities define the additional intersection numbers 
\ba
M_{i a}{}^\k & = \int_{Y_3} \g_i \we \o_a \we \b^\k \, , & 
N_{ij}{}^\a &= \int_{Y_3} \tilde \omega^\alpha \wedge \gamma_i \wedge \gamma_j \, . 
\ea
Compatibility of these conditions then implies that 
\ba
\label{kMConditionsFromIIB}
k_{i}^a M_{j a}{}^\k & =  k_{j}^a M_{i a}{}^\k \, , &  
k_{\a \k} M_{i a}{}^\k & = k_i^b \cK_{\a a b} \, ,   & 
k_{\a \k} N_{i j}{}^\a &= 0  \, . 
\ea
In the second identity in (\ref{complete}), we have allowed for a non-trivial product between the torsion two-forms which, as we will see, is coupling responsible for a non-Abelian gauge symmetry.

Given this setup, the ansatz for the reduction is
\bea \label{C4C2expansion}
    C_4 &=& V^\kappa \wedge \alpha_\kappa - U_\kappa \wedge \beta^\kappa + \rho_\alpha \tilde \omega^\alpha + C_2^\alpha \wedge \omega_\alpha\ ,\\[.1cm]
    B_2 &=& A^{1\, i} \wedge \gamma_i +  b^a \omega_a\ , \qquad C_2  = A^{2\, i} \wedge \gamma_i +  c^a \omega_a \ . \nn 
\eea
where $C_4$ has an expansion into orientifold-even three-forms while $B_2,C_2$ are expanded into orientifold-odd one-forms and two-forms.
Here $A^{1\, i}, A^{2\, i}$ and $V^\kappa,U_{\kappa}$ are four-dimensional vectors.
Note that $V^\kappa$ and $U_{\kappa}$ are electric-magnetic 
duals by means of the self-duality of the field-strength of $C_4$. Similarly,
the two-form $C_2^\alpha$ is the four-dimensional dual of the scalar 
$\rho_\alpha$ already used in \eqref{schem_text}. 

The effective action which results from the ansatz  \eqref{C4C2expansion} can be described in terms of the fields $C_2^\alpha$ and $U_\kappa$ or in terms of their duals  $\rho_\alpha$ and $V^\kappa$. When working with 
$C_2^\alpha$ and $U_\kappa$ the 10d field strength 
\ba
F_5 = dC_4 + \frac12 (B_2 \wedge dC_2 - C_2 \wedge dB_2) \, , 
\ea
gives rise to the four-dimensional field strengths 
\ba 
  DC_2^\alpha &= dC_2^\alpha + \frac{1}{2} N^\alpha_{ij}(A^{1\, i} \wedge F^{2\, j} - A^{2\, i} \wedge F^{1\, j}) \,, & 
   F_{\kappa} &= dU_{\kappa} - k_{\alpha \kappa} C_2^\alpha \ , 
\ea
where
\ba
  F^{1\, i} &= d A^{1\, i}\ , &
  F^{2\, i} &= d A^{2\, i}\ , 
\ea
Here we see that the nonlinear terms in $F_5$ have generated a Chern-Simons modification in $DC_2^\a$, but that all gaugings remain Abelian.

In contrast, if one works in the dual picture and encodes all degrees 
of freedom by $\rho_\alpha$ and $V^\kappa$, one finds the 
field strengths 
\ba
   F^{1\, i} &= d A^{1\, i} \,  ,& 
   F^{2\, i} &= d A^{2\, i}\, , &
   F^\kappa &= d V^\kappa +   M_{i a}{}^\k k^{a}_j A^{1\, i} \wedge A^{2\, j}\, ,
\ea
where, in this dual picture, the nonlinear terms in  $F_5$ have generated a non-Abelian structure $F^\k$. In fact, one checks by performing
the reduction that the isometries of \eqref{GTisom} are 
gauged due to the non-trivial $k_{\alpha \kappa}$ and 
$k^a_i$. Explicitly, the covariant derivatives read
\ba 
\label{con_TG}
   DT_{\alpha} &= dT_\alpha  + k_{\alpha \kappa} V^\kappa  -  \cK_{\alpha a b} G^a k_{i}^b A^{1\, i} \ , & 
   DG^a &= dG^a +  k^a_i (A^{2\, i} -\tau A^{1\, i} ) \ .
\ea
This suggests that the gaugings are compatible with the holomorphic 
structure of the reductions. However, by performing the dimensional reduction \cite{BerasaluceGonzalez:2012vb} we see that the gauge coupling function derived fails to be holomorphic in the coordinates introduced above. We therefore propose that this construction is not compatible with supersymmetry without modifying the ansatz \eqref{domega_ori} and \eqref{C4C2expansion}.

Let us add some more observations to support this further. We stress 
that there is a curiosity in the gaugings \eqref{con_TG}: for the gauged scalars $G^a$ 
it appears that the real and imaginary parts are gauged \emph{at the same time} 
with two different vectors corresponding to non-commuting generators. We find that the K\"ahler potential both depends on these scalars and is invariant under the symmetry. This property of the gauged $G^a$ implies that constructing a holomorphic gauge coupling function which transforms in the appropriate fashion appears to be impossible.
Furthermore we see that 
in the underlying $\cN=2$ theory obtained by a Calabi-Yau reduction the fields 
completing $c^a,b^a$ into hypermulitpletshypermultiplets are the scalars $\rho_a$ from $C_4$ and 
$v^a$ from the K\"ahler form. One can check that these scalars are ungauged 
and admit a scalar potential. This shows that the truncation associated with the orientifold quotient inconsistently removes the two ungauged degrees of freedom from the hypermultiplets.

We therefore find that the inclusion of  torsion cohomology is by no means straightforward in the presence of an orientifold projection. It would be interesting to reveal the underlying physical reason of the incompatibility of the $\cN=1$ orientifold truncation with the torsion proposal of \cite{BerasaluceGonzalez:2012vb}.  Our findings suggest that  torsion cohomology can only be `straightforwardly' included for orientifold-even forms, i.e.~where there are chains associated to the forms with non-vanishing physical volume. In the next subsection we will argue that when generalising the setting away form the weak string coupling limit the gaugings can be made compatible with $\cN=1$ supersymmetry.

\section{Non-Abelian discrete symmetries in F-theory} \label{nonA_inFtheory}

In this section we discuss the appearance of four-dimensional
non-Abelian discrete symmetries in brane-bulk and brane-flux systems from the Type IIB perspective. 
This will allow us to develop different settings that 
naturally admit such symmetries. In order to realise these symmetries in a seven-brane 
sector, however, it that this requires the introduction of mutually non-local branes that 
cannot be treated at weak string coupling. To find a globally consistent 
description of such a system we therefore will use F-theory. While we 
are able to heuristically motivate our findings directly using the language of Type IIB 
string theory with $(p,q)$-seven-branes a more thorough justification 
will later, in section \ref{MFduality}, be given by using the M-theory approach to F-theory.

\subsection{Heisenberg symmetries in non-perturbative Type IIB} \label{non-pert_TypeIIB}

In the previous subsection we have shown that the non-harmonic 
reduction yielding a non-Abelian gauge theory is not compatible 
with the $\cN=1$ supersymmetry imposed by the Type IIB orientifold 
projection. This conflict arose from the fact that the modes arising from the fields 
$B_2$ and $C_2$ naturally combine into $\cN=1$ four-dimensional 
scalars $G^a = c^a -\tau b^a$
with real and imaginary parts simultaneously gauged by two different 
vector fields.  
Importantly, this analysis was a weak string coupling 
analysis in which the ten-dimensional $\tau$ did not
vary over the internal manifold but rather descended to a four-dimensional degree of freedom. This 
led to the fact that the behaviour of the modes $b^a$ and $c^a$ 
cannot be decoupled. However, in the more general situation in which we depart 
from weak string coupling, the $Sl(2,\mathbb{Z})$ symmetry group 
will have non-trivial monodromies on the compactification space 
and neither $\tau$ nor $c^a,b^a$ are well-defined fields in the 
effective theory. In the following we introduce the analogs for the 
fields $G^a$ in compactifications with varying $\tau$ and 
describe how the coupling to seven-branes introduces a 
non-Abelian gauge structure. 

Let us now work on the K\"ahler manifold $\cB_3$, which 
is the base of an elliptically fibered Calabi-Yau fourfold $Y_4$ that we 
use for the F-theory treatment. In 
settings with weak coupling limit one has $\cB_3 = Y_3/\mathbb{Z}_2$.
When working in Type IIB language one would need to expand 
\beq \label{generalized_expansion}
   C_2 - \tau B_2  =    N^a \hat \Psi_a\ ,
\eeq
where $\hat\Psi_a$ are appropriate two-forms on an $Sl(2,\mathbb{Z})$-bundle on $\cB_3$ and
$ N^a$ are complex scalar fields in four dimensions. The two-forms $\hat \Psi_a$ 
will in general depend on the complex structure moduli of $\cB_3$ and the seven-brane positions. 
It is expected that the explicit construction of the $\hat\Psi_a$ is challenging. 
However, the Calabi-Yau fourfold $Y_4$ turns out to be a powerful tool to capture this information 
in a more tractable way. 

On $Y_4$ the information encoded in $\hat \Psi_a$
is captured by a certain basis of (2,1)-forms $\Psi_a$ that do not descend from $(2,1)$-forms of the base $\cB_3$. 
The additional constraint is often stated as the requirement that the $\Psi_a$ have a component with one leg 
in the fiber of $Y_4$.   
In the simplest situation $\Psi_a$ are harmonic forms that are parameterising $H^{2,1}(Y_4)$ but 
are not elements of $H^{2,1}(\cB_3)$. In the following, we will first consider only harmonic $(2,1)$-forms,
but later generalize to include non-closed and exact forms. 
To obtain the four-dimensional fields $N^a$ in \eqref{generalized_expansion}
one now has to expand a three-form $C_3$ into the $(2,1)$-forms $\Psi_a$ as follows
\be \label{expand_C3complex}
C_3= N^a \Psi_a+\bar N^a\bar\Psi_a+ \dots \ .
\ee
This is motivated by the M-theory to F-theory limit as we discuss in section \ref{MFduality}. 
In this limit the $N^a$ lift to four-dimensional scalars that include the scalars coming from $B_2, C_2$.
Furthermore, despite the abuse of notation for the indices, the $N^a$ will 
also contain the seven-brane Wilson lines. 
To display the effective action one first has to gain some deeper understanding of the moduli dependence 
of $\Psi_a$. Clearly, since these are $(2,1)$-forms on $Y_4$ they will vary with the complex structure moduli 
of $Y_4$. For $H^{2,1}(Y_4)$ one can in fact argue that 
the $\Psi_a$ admit an expansion
\ba
\label{def-Psif}
\Psi_a & =\frac{1}{2}{\rm Re}f_{ab}(\b^b-i\bar f^{bc}\a_c)\, , &
 \Psi_a-\bar\Psi_a &=-i\a_a \,, 
\ea
where $(\a_a,\b^a)$ are a real three-form basis for the elements of $H^3(Y_4)$ 
which are not in $H^{2,1}(\cB_3)$ and $f^{ab}$ is a holomorphic function of the complex structure moduli. 
The $\Psi_a$ are not anti-holomorphic in the complex structure moduli 
due to the appearance of ${\rm Re} f_{ab}$, the inverse of ${\rm Re}f^{ab}$. 
Using the real basis $(\a_a,\b^a)$ we can also expand 
\be \label{expand_C3real}
C_3=a^a\a_a-b_a\b^a+\dots
\ee
where $(a^a,b_a)$ are real scalars. Comparing \eqref{expand_C3complex} with \eqref{expand_C3real} and using \eqref{def-Psif}, we see that 
\be \label{def-Ncoords}
N^a=-i(a^a+i f^{ab}b_b)\ .
\ee

In a next step we recall the effective action for the complex scalars $N^a$ coupled
to $v^\alpha,\rho_\alpha$ and study its 
symmetries. The derivation of this action proceeds via M-theory as carried out in \cite{Grimm:2010ks}.
This yields the generalisation of the weak string coupling action \eqref{schem_text}  to F-theory as
\bea \label{F-theoryaction}
\mathcal L&=&-G_{\a\b}dv^\a\w* dv^\b-\frac{1}{4 \mathcal V^2}d \mathcal V\w*d\mathcal V
+ \frac{3v^\a}{\mathcal K}d_{\a ab}\,dN^a\w *d\bar N^b\\ \nonumber
&&-\frac{G^{\a\b}}{16 \mathcal V^2}\big(d\rho_\a+\tfrac{i}{2}(d_{\a ac}\bar N^cdN^a-d_{\a ca}N^cd\bar N^a)\big)\! \w\! *\big(d\rho_\b+\tfrac{i}{2}(d_{\b bd }\bar N^ddN^b-d_{\b db }N^dd\bar N^b)\big)\, .
\eea
with
\be
d_{\a ab}=i\int_{Y_4}\omega_\a\,\w\, \Psi_a\,\w\, \bar \Psi_b\ .
\ee
Here $\omega_\a$ is a two-form dual to vertical divisors $D_\a=\pi^{-1}(D_\a^{\rm b})$,
where $D_\a^{\rm b}$ are divisors in the base $\cB_3$. 
Thus, $\bar d_{\a ab}=d_{\a ba}$. In terms of the real basis we have that
\be
d_{\a ab}=\frac{1}{2}{\rm Re}f_{ac}\left ( M_{\a b}{}^{c}+i\bar f^{cd}M_{\a d b}\right )\ ,
\ee
%d_{\a ab}=\frac{i}{4}{\rm Re} f_{ac}{\rm Re}f_{be}\left ( M_{\a}{}^{ce}+\bar f^{eg}f^{cd}M_{\a dg}-if^{eg}M_{\a g}{}^{c}-i\bar f^{cd}M_{\a d}{}^e\right )
where we defined
\be
M_{\a ab}=\int \omega_\a\wedge \a_a\wedge \a_b,\qquad \quad M_{\a a}{}^b=\int \omega_\a\wedge \a_a\wedge \b^b\ .
\ee
%\quad M_{\a}{}^{ab}=\int \omega_\a\wedge \b^a\wedge \b^b. 

The action \eqref{F-theoryaction} can be expressed in terms of a K\"ahler potential and 
complex coordinates as in the weak string coupling setting. The correct complex
coordinates are the $N^a$ as well as complex coordinates $T_\a$ containing the 
K\"ahler structure deformation defined as
\be \label{def-Talpah_F}
T_\a=\rho_\a-\frac{i}{2}d_{\a ab}N^a(N+\bar N)^b-\frac{i}{2}\int_{D_\a^{\rm b}}J_{\rm b}\wedge J_{\rm b}
\ee
where $J_{\rm b}$ is the K\"ahler form in the base. The K\"ahler potential is 
given by 
\beq
   K = - \text{log} \Big( \int_{Y_4} \Omega \wedge \bar \Omega \Big) - 2 \text{log} \mathcal{V}_{\rm b}\ , 
\eeq
where it is crucial to express the base volume $ \mathcal{V}_{\rm b} = \frac{1}{6} \int_{\cB_3} J_{\rm b} \wedge J_{\rm b} \wedge J_{\rm b}$ in terms of the complex coordinates $N^a$, $T_\alpha$ given in \eqref{def-Talpah_F}, 
and the complex structure deformations.

Let us now turn to the analysis of the isometries of the metric \eqref{F-theoryaction}.
The metric has the following holomorphic isometries
\ba
\label{isomF}
\delta N^a=&\,\,-i(\lam^a+i f^{ab}\lam_b) \,, \nn \\
\delta T_\a=&\,\,\lam_\a-\frac{i}{2}N^b M_{\a ab}\lam^a-N^a \Big(iM_{\a a}{}^b+\frac{1}{2}f^{bc}M_{\a ca}\Big)\lam_b \, , 
\ea
%\d T_\a=&\,\,\lam_\a-iN^b\left [{\rm Im}( d_{\a ab})\lam^a+({\rm Re}(f^{ac})d_{\a ba}+{\rm Re}(f^{ac}d_{\a ab}))\lam_c\right ]
with $\lam^a, \lam_a,\lam_\a$ real. The corresponding Killing vectors read
\bea 
\label{FtGens} 
\til t^b&=& \,\,f^{ab}\p_{N^a}-N^a \left (iM_{\a a}{}^b+\frac{1}{2}f^{bc}M_{\a ca}\right )\p_{T_\a} \ ,\\
 t_a&=&\,\,-i\p_{N^a}-\frac{i}{2}N^b M_{\a ab}\p_{T_\a}\ ,\qquad 
  t^\a=\,\,\p_{T_\a}\ .
\eea
It is then straightforward to check that the only non-vanishing commutator is
\be \label{HeissenbergF}
[t_a, \til t^b]=-M_{\a a}{}^b\, t^\a\ ,
\ee
which again defines an algebra that is a generalisation of the Heisenberg algebra. 
Notice that $M_{\a ab}$ does \emph{not} appear in the structure constants.

The expression \eqref{HeissenbergF} is the analog of the weak string coupling algebra \eqref{Heisenberg_complex}. In 
fact, the setup reduces to the one of subsection \ref{Heissenberg_ori} in a special limit.
In order to see that one interprets all fields $N^a$ to arise from the bulk as 
the fields $G^a$ used in subsection \ref{Heissenberg_ori}.
Setting 
\ba \label{fN_weak}
   f^{ab} &= i \tau \delta^{ab} \ , &
   N^a  &= -  i G^a \ , 
\ea
one recovers the weak coupling expressions for all couplings.
However, it is crucial to point out that away from weak string coupling $f^{ab}$ will in general not 
be diagonal. The non-diagonal generalisation 
will be crucial when considering the gauging of the holomorphic isometries as 
we discuss in the next subsection. In contrast to the weak coupling setting there can now be gauged scalars $N^a$ for which the real and imaginary parts are not gauged simultaneously while preserving the non-Abelian structure.

It is interesting to stress that in F-theory the $N^a$ also contain the Wilson line degrees of 
freedom. Even at weak string coupling, i.e.~when considering $N^a$ to be Wilson line 
moduli for D7-branes, one finds that they couple via a holomorphic function $f^{ab}$
of the complex structure moduli and D7-brane positions. 
It appears that this holomorphic function does not have to be diagonal in its indices. 
This yields another non-trivial generalisation 
of the setting discussed in subsection \ref{Heissenberg_ori}. In F-theory the various generalisations 
are elegantly combined due to the combination of bulk and brane degrees of freedom in 
a higher-dimensional geometry.

\subsection{Non-Abelian gaugings from seven-branes - Origins} \label{sec:origins}

Having determined the holomorphic symmetries of the general Type IIB configuration 
away from the weak string coupling limit, one can now ask which subalgebra of these symmetries can be gauged. In particular, given the complications encountered 
for the orientifold setup in subsection \ref{susygaugings_ori}, it will be crucial to argue that in more 
general F-theory settings a non-Abelian group can indeed be gauged. The gaugings 
we will discuss arise from gauge fields on general $(p,q)$-seven-branes and we  
also consider possible gaugings using R-R gauge fields due to non-closed forms
in the base $\cB_3$. The non-closed 
forms can be interpreted as parameterising torsional cohomology $\text{Tor} H^3(\cB_3,\mathbb{Z})$
similar to the discussion of subsection \ref{susygaugings_ori}.

To begin with we first recall the gaugings arising when  D7-branes are included
in a weak string coupling scenario. 
If we include D7-branes wrapping a divisor $S_i$, the $U(1)$ gauge boson $A^i$ with field-strength $F^i=dA^i$ may become 
massive due to the interaction with the closed string sector for two independent reasons, which in either case are compatible with supersymmetry. 
 
Firstly, we consider a configuration with brane image-brane pairs in an orientifold configuration 
in which some of the divisors $S_i$ and the image-brane divisors $S'_i$ are in different homology classes,
i.e.~situations in which some of the $S^-_i=\frac{1}{2}(S_i-S_i')$ are homologically non-trivial. 
 The Chern-Simons action then contains  
 a coupling of the form %\footnote{We do not consider the possibility of flux-induced St\"uckelberg couplings.}
\ba
\label{csd7}
S_{\text{D7}_i} & \supset \int_{\mathbb R^{1,3}\times S^-_i}C_6\wedge F^i = \til k^a_i \int_{\mathbb R^{1,3}}\tilde c_a^{(2)}\wedge F^i \, , &
 \til k^a_i &=  \int_{S^-_i}\tilde\omega^a \ 
\ea
where we have expanded $C_6=\tilde c_a^{(2)}\wedge \tilde \omega^a$ with  $\tilde\omega^a$ being an integral harmonic four-form that is odd under the orientifold action. This induces a St\"uckelberg gauging of the axion $c^a$ dual to the two-form $\tilde c_a^{(2)}$ as
\ba
 \label{geom_stueck}
    DG^a &= d G^a - \til k^a_i A^i\  , 
\ea
One can show that, generically, this leads to a spontaneous breaking of the symmetry (see e.g. \cite{Banks:2010zn}). The surviving unbroken symmetry my contain a discrete part which is always \emph{Abelian}. The details of this discrete part are discussed in \cite{BerasaluceGonzalez:2011wy,BerasaluceGonzalez:2012zn}.

Secondly, there is a possibility of switching on fluxes $\mathcal{F}^{i}$ on the D7-branes. 
The gauging induced by this generalisation is of the form
\ba
\label{flux_stueck}
     D T_\alpha &= dT_\alpha - \Theta_{\alpha i}  A^i \ , &
     \Theta_{\alpha i } &=   \int_{S_i} \mathcal{F}^{i} \wedge \omega_\alpha\ . 
\ea
with $DG^a$ being unmodified. We note that these
considerations generalize if we include several D7-branes. 
Taking into account the appropriate Chern-Simons couplings, we may find a discrete \emph{Abelian} gauge symmetry \cite{BerasaluceGonzalez:2011wy,BerasaluceGonzalez:2012zn}. 

To gain a intuition how this D7-brane setting generalises, let us naively
consider a configuration that contains 
O7-planes and (0,1)-seven-branes. In this case, the analogous coupling to (\ref{csd7}) is
\begin{equation} \label{01Stueck}
S_{(0,1)}\supset \int_{\mathbb R^{1,3}\times S^-_i}B_6\wedge F^i =  \delta^{ab} k_{b i}\int_{\mathbb R^{1,3}}\tilde b_a^{(2)}\wedge F^i,
\end{equation}
with $B_6$ dual to the NS-NS two-form, $B_6=\tilde b_a^{(2)}\wedge\tilde \omega^a$. 
One would therefore expect that in this case the $b^a$ scalar, dual to $\tilde b_a^{(2)}$, receives a gauging of the form
\beq
\label{01_stueckk}
   D G^a = d G^a - \tau \delta^{ab} k_{a i} A^i\ . 
\eeq
Of course, this setting cannot be fully trusted, since we have included a $(0,1)$-seven-brane in a weak coupling 
scenario. We should instead return to the F-theory setting outlined in subsection \ref{non-pert_TypeIIB} as we will do below.

Let us finally turn to the discussion of gaugings due to non-closed two-forms 
in the base $\cB_3$. This will lead to gaugings involving the R-R gauge-fields 
just as in subsection \ref{susygaugings_ori}. 
As before this requires non-closed forms to be included among the 
$\omega_\alpha$ in the base $\cB_3$ such that 
\beq \label{domega_base}
    d \omega_\alpha  =  k_{\alpha \kappa}  \beta^\kappa \ ,
\eeq
 where $\beta^\kappa$ are three-forms in $\cB_3$. 
 Carrying out the expansion of $C_4$ in a process similar so that shown in subsection \ref{susygaugings_ori} one finds that \eqref{domega_base} 
 induces the gauging 
 \ba
\label{tor_stueck}
     D T_\alpha &= dT_\alpha - k_{\alpha \k}  A^\k \ ,  
\ea
which is of similar form as \eqref{flux_stueck} but only uses the bulk vectors $A^\k$.
 The relation \eqref{domega_base} 
can be interpreted as arising from torsional cohomology Tor$H^3(\cB_3,\mathbb{Z}) \cong \text{Tor}H^4(\cB_3,\mathbb{Z})$
as introduced in subsection  \ref{susygaugings_ori}. Note that the $H^p(\cB_3)$ have to be
identified with $H^3_{+}(Y_3)$ if a double-covering Calabi-Yau threefold $Y_3$ exists in 
the weak coupling limit. We thus do not require that torsion in the negative cohomology be considered. This modification of the setting may evade the problems encountered in subsection \ref{susygaugings_ori}.

\subsection{Non-Abelian gaugings from seven-branes - Gauge invariant structures} \label{sec:invariant_structures}

We have just motivated that the gaugings in a Type IIB setting with $(p,q)$-seven-branes can
be more general than in the weak coupling configurations of section \ref{susygaugings_ori}.
In order to study the system away from the weak string coupling limit we return to 
the configuration introduced in subsection \ref{non-pert_TypeIIB}.
To gain some intuition about the gaugings that occur
one can formally perform the replacement \eqref{fN_weak} introducing $N^a$ and $f^{ab}$
in the gaugings of subsection \ref{sec:origins}. 
An honest derivation, however, can only be performed via the duality with M-theory. 
In fact we will justify some of the following results using this duality in section \ref{MFduality}.

In general, one finds that only a subalgebra of the isometry algebra \eqref{isomF} discussed in 
subsection \ref{non-pert_TypeIIB} will be gauged. Clearly, to define a subalgebra 
one has to respect various constraints ensuring, for example, the closure of this 
algebra. The structure constants will generically also differ from the ones 
of the full isometry algebra. Let us exemplify this by using a subset 
of the gaugings introduced in subsection \ref{sec:origins}.
In a first F-theory example will only use seven-brane vectors in the gaugings,
and hence the structure constants of the subalgebra will be of the form $\hat f_{i j}{}^k$.
Motivated by the structures which appear in \eqref{geom_stueck}, \eqref{flux_stueck} and \eqref{01_stueckk} we will consider a subalgebra of \eqref{isomF} that is associated with the generators 
\ba
\label{SubAlgebraDef}
X_i & = k_i^a  t_a - k_{ia} \til t^a + \Q_{ \a i}  t^\a \, , &
[t_i ,t_j]& = \hat f_{i j}{}^k t_k \, ,
\ea
which defines the structure constants $ \hat f_{i j}{}^k $.
Then by using \eqref{HeissenbergF} we find that
\ba
\label{fMkConstraintsFlux}
 (\tilde k^a_i k_{j b} - \tilde k^a_j k_{i b}) M_{\alpha a}{}^b & = \hat f_{i j}{}^k \Q_{\a k}\, , & 
 \hat f_{i j}{}^k k_{k a} & = 0 \, , & 
  \hat f_{i j}{}^k \til k_{ k}^a & = 0 \, .  
\ea
We note that this analysis is not sufficient to uniquely fix the structure constants $ \hat f_{i j}{}^k$ but only certain projections on them. This is familiar from the standard embedding tensor discussions (see e.g.~\cite{Samtleben:2008pe,deWit:2004yr}). 
The covariant derivatives associated with gauging this subgroup are then given by 
\bea \label{cov-derv-Ftheory}
DN^a&=& dN^a + i(\tilde k^a_i A^i-i f^{ab} k_{ib} A^i)\ ,\\
DT_\a&=&dT_{\alpha} - \Theta_{\alpha i} A^i + \frac{i}{2}N^b M_{\a ab} \tilde k^a_i A^i - N^a \Big(iM_{\a a}{}^b+\frac{1}{2}f^{bc}M_{\a ca}\Big) k_{ib}A^i\ , \nn 
\eea
and the field strength $F^i = dA^i + \hat f_{j k}{}^i A ^j \we A^k$ is constrained such that 
\ba
\label{fieldstrength_braneFtheory}
\Q_{\a i} F^i & = \Q_{\a i}  d A^i +   \tilde k^a_k k_{j b}M_{\alpha a}{}^b A^j \we A^k \, , & 
k_{i a} F^i & = k_{i a} d A^i \,  ,&
k_{i}^a F^i & = k_{i}^a d A^i \, . 
\ea
If we recall that, roughly speaking, $\tilde k^a_j $ labels $(1,0)$-brane part of the gauging, $k_{jb}$ is the $(0,1)$-brane part of the gauging, and $M_{\alpha a}{}^b $ is the non-trivial twisting of the moduli space metric \eqref{F-theoryaction}, then we see that it is the presence of gaugings associated with mutually non-local seven-branes that is crucial for generating the non-Abelian structure in \eqref{fieldstrength_braneFtheory}.
In addition to this we see that the non-Abelian structure is linked to the presence of 
fluxes in this picture. It is well-known that fluxes induce chirality and accordingly the classical action does not 
need to be gauge invariant as discussed briefly at the end of subsection \ref{non-min-kinetic-terms}. 

The second example of non-Abelian gaugings occurring in F-theory is obtained by
switching off fluxes on the seven-branes (i.e.~setting $\Theta_{\alpha i}=0$) and turning on $k_{\a \k}$
appearing in \eqref{domega_base}. It will be this example that we will study in much more detail using the 
M-theory dual in section \ref{MFduality}. Analysing the subalgebra spanned by $\tilde k^a_j,\, k_{i a}$, and $k_{\a \k}$, we
find that the only non-vanishing structure constants are in this case of the form 
$\hat f_{ij}{}^\k$. They are constrained only by  
\ba
\label{GaugedStructrueConstantsRR}
 \hat f_{i j}{}^\k  \Pi_\k{}^\l &= (\tilde k^a_i k_{j b} - \tilde k^a_j k_{i b}) M_{\alpha a}{}^b \check k^{ \l\a } \, , 
\ea
and that for the gauged subalgebra to close we must demand that
\ba
\Pi_\a{}^\b  (\tilde k^a_j k_{i b} - \tilde k^a_i k_{j b}) M_{\b a}{}^b  & = (\tilde k^a_j k_{i b} - \tilde k^a_i k_{j b}) M_{\a a}{}^b \, . 
\ea
In these equations we have defined the projectors $ \Pi_\k{}^\l$ and $ \Pi_\a{}^\b$ as well as the Moore-Penrose pseudo-inverse $\check k^{ \k\a }$ of $k_{\a \k}$. These quantities satisfy 
\ba
k_{\a \k }  \check k^{\l \a} & = \Pi_\k{}^\l \, , & 
k_{\a \k } \check k^{\k \b} & = \Pi_\a{}^\b \, , & 
\Pi_\k{}^\l k_{ \a \l}& = k_{ \a \k } \, ,&
\Pi_\a{}^\b k_{\b \k} & = k_{ \a \k} \, . 
\ea
In this case the gaugings \eqref{cov-derv-Ftheory} are replaced by 
\ba \label{cov-derv-Ftheory_noncl}
DN^a&= dN^a+i(\tilde k^a_i A^i-i f^{ab} k_{ib} A^i)\ , \nn \\
DT_\a&=dT_{\alpha}- k_{\alpha \kappa} A^\kappa +\frac{i}{2}N^b M_{\a ab} \tilde k^a_i A^i -N^a \Big(iM_{\a a}{}^b+\frac{1}{2}f^{bc}M_{\a ca}\Big) k_{bi}A^i\ \, , 
\ea
and the field strengths are constrained such that
\ba\label{fieldstrengths}
k_{\a \k} F^\k&= k_{\a \k}dA^\k+  \tilde k^a_j k_{k b} M_{\alpha a}{}^b A^j \we A^k  \, , & 
F^i = dA^i\ . 
\ea
We stress that in this case only the R-R bulk gauge-field admits a non-Abelian modification. 

This second possibility of obtaining non-Abelian gaugings has the advantage of being purely 
geometrically induced. In particular, one expects following \cite{Grimm:2010ez,Grimm:2011tb,Grimm:2013fua,Braun:2014nva} that 
the geometrically massive gauge fields gauging $N^a$ are obtained from non-closed 
forms on the Calabi-Yau fourfold in M-theory. Together with the possibly non-closed 
two-forms $\omega_\alpha$  satisfying \eqref{domega_base}, one thus expects to 
find a geometric M-theory reduction that yields precisely the gaugings \eqref{cov-derv-Ftheory_noncl}
upon lifting to F-theory. We will show in section \ref{MFduality} that this is indeed the case. 
Furthermore, we are able to directly determine the structure constants $\hat f_{ij}{}^{\k}$
to be given by 
\beq \label{structure_mixed}
\hat f_{ij}{}^{\k} = \tilde k_{ [j}^a M^{\phantom{a}}_{i] a}{}^\k+ k_{[ja}M_{i]}{}^{a\k}\ .
\eeq
Here $M^{\phantom{a}}_{i a}{}^\k$ and $M_{i}{}^{a\k}$ are constant coupling 
matrices that are explicitly given in section~\ref{MFduality}. 

To get an idea about the meaning of these couplings, let us give their 
weak string coupling expressions in the Calabi-Yau orientifold setting $\cB_3 = Y_3 /\sigma$. 
If the index $a$ counts bulk scalars $G^a$ then we find for D7-branes
\beq \label{Mori}
    M^{\phantom{a}}_{i a}{}^\k = 0 \ , \qquad M_{i}{}^{a\k} = \delta^{ab} \int_{\cC^i} \omega_b \wedge \beta^\k\ .
\eeq
where $\cC^i$ is a chain ending on the $i$th D7-brane world-volume and $\omega_a$ is 
the orientifold-odd harmonic two-form on $Y_3$. Note that $M^{\phantom{a}}_{i a}{}^\k$, 
as defined in section~\ref{MFduality}, should only include the constant part of the 
chain integral in \eqref{Mori}. Once again we can see that the non-Abelian gaugings disappear 
for settings with only D7-branes, since in this case $k_{ia}=0$ and $ M^{\phantom{a}}_{i a}{}^\k = 0$
in \eqref{structure_mixed}.

Alternatively the index $a$ could also parameterise Wilson line moduli on the seven-branes.
Let us introduce one-forms $(\gamma_{a_i},\gamma^{b_i})$ on the $i$th seven-brane
with world-volume $S^i$. Then we find that 
\beq \label{MWL}
     M^{\phantom{a}}_{i a_i}{}^\k = \int_{S^i} \gamma_{a_i} \wedge \beta^\k \ , \qquad 
     M_{i}{}^{a_i \k} = \int_{S^i} \gamma^{a_i} \wedge \beta^\k\ .
\eeq
In this case one finds indeed that both $M^{\phantom{a}}_{i a_i}{}^\k$ and $M_{i}{}^{a_i \k}  $
are non-zero. However, in order to realise a non-Abelian symmetry with 
non-vanishing \eqref{structure_mixed} we need to gauge the Wilson line scalars. 
We are not aware that such a setting has been investigated yet. 

Let us close this section with another crucial observation which ties in with 
the discussion of the gauge coupling function presented at the end of 
subsection \ref{non-min-kinetic-terms}. It also explains 
how we were able to deduce the expressions \eqref{Mori} and \eqref{MWL}.
It turns out, as we will see in section \ref{MFduality}, that 
the $M^{\phantom{a}}_{i a}{}^\k$ and $M_{i}{}^{a\k}$ precisely
encode the kinetic mixing of the R-R gauge fields $A^\k$ and the 
seven-brane gauge-fields. 
More precisely, we find 
\bea \label{gaugecoupling_F}
     \R f_{\l i} &=& \R f_{\l \k}(M_{ ia}{}^\k a^a-M_{ i}{}^{a\k} b_a)\ , \\
     \R f_{ij} &=& \check G_{ i j}  + \R f_{\l \k}  (M_{ ia}{}^\k a^a-M_{ i}{}^{a\k} b_a)(M_{ ja}{}^\l a^a-M_{ j}{}^{a\l} b_a)   \ , \nn 
\eea
where $f_{\l \k}$ is the holomorphic gauge coupling function of the R-R gauge fields $A^\k$, 
and $ \check G_{ i j} $ is a K\"ahler moduli dependent metric. The fact that 
the gauge couplings depend on the scalars $(a^a,b_a)$ nicely 
matches the requirement that for a gauged non-semisimple and non-compact
group this coupling needs to transform non-trivially. The kinetic mixing \eqref{gaugecoupling_F}
is present independent of the gaugings, i.e.~even if we set $\tilde k^i_a=k_{ia}=0$ and $k_{\alpha \k}=0$. 
If we allow for non-Abelian gaugings than the terms in \eqref{gaugecoupling_F} are actually essential
for gauge-invariance. Let us note that the results  \eqref{Mori} and \eqref{MWL} were deduced 
by comparing the kinetic mixing terms on seven-branes with \eqref{gaugecoupling_F}.
Kinetic mixing on D7-branes was studied also in \cite{Jockers:2004yj,Abel:2008ai,Marchesano:2014bia}. 
One therefore expects that \eqref{gaugecoupling_F} can be made a real part of a
holomorphic function in the correct $\cN=1$ complex coordinates as required by supersymmetry. 
We leave the details of this investigation to a further publication \cite{toAppearKinMixing}.

Let us close this section by stressing some of the differences to the 
discrete Abelian symmetries recently considered in \cite{Braun:2014oya,Morrison:2014era,Anderson:2014yva,Klevers:2014bqa,Garcia-Etxebarria:2014qua,Mayrhofer:2014haa, Mayrhofer:2014laa,Cvetic:2015moa}. 
As of now, most of the considerations were for the effective theory and 
the continuous non-Abelian symmetry. Focusing on the vacua of the theory
one expects that there is, in contrast to the Abelian case, no vacuum in which 
a continuous non-Abelian group is preserved. This can be inferred from the fact
that the background gauge coupling function can not be positive definite and invariant as no such tensor exists. 
In the Abelian case
a more complete analysis was possible and it was argued that in this 
case there exists a transition in 
complex structure moduli space that restores a global U(1) symmetry. 

\section{Non-Abelian discrete symmetries via F-/M-theory duality} \label{MFduality}

In this section we use the duality of M-theory and F-theory to show 
the appearance of discrete non-Abelian gauge symmetries in F-theory as 
claimed in section \ref{nonA_inFtheory}. 
More precisely, we perform a dimensional reduction 
of eleven-dimensional supergravity including a number of non-harmonic forms. These
forms might be viewed as representing torsion cohomology elements. The 
three-dimensional effective action is determined in subsection \ref{nonharm_Mreduction}
and possesses only Abelian gaugings.
The non-Abelian gaugings arise when bringing the three-dimensional action into
the duality frame relevant for the F-theory up-lift to four dimensions. The relevant 
dualisations of the fields are discussed in subsection \ref{dualizing_action} and appendix \ref{DualAction}.
We are then able to show that the covariant derivatives \eqref{cov-derv-Ftheory_noncl} and field strengths \eqref{fieldstrengths} are reproduced by the reduction. 
We also find that the structure constants are given by \eqref{structure_mixed} and the gauge coupling function takes the form \eqref{gaugecoupling_F}.

\subsection{Non-harmonic reduction of M-theory} \label{nonharm_Mreduction}

Recall that the duality between M-theory and F-theory asserts that compactifying 
the former theory on an elliptically fibered Calabi-Yau manifold is dual 
to the latter theory on the same manifold times a circle. The comparison 
of effective theories of M-theory and F-theory is therefore performed in three
dimensions. One can thus start with a candidate four-dimensional action, the F-theory 
effective action, and compactify the theory on a circle. The lower-dimensional
theory can be pushed to the Coulomb branch and all heavy modes, including the Kaluza-Klein states,
can be integrated out to obtain the effective theory for massless states only. 
However, we claim that the M-theory reduction will also contain massive modes
that arise due to the inclusion of non-harmonic forms. Therefore, we have 
to carefully keep track of certain charged or massive states in the 
matching of the M-theory and F-theory actions. This is in complete analogy 
to the case in which one considers background fluxes. In the following we will thus 
discuss three-dimensional gauged supergravity theories to justify 
the F-theory effective action of section \ref{nonA_inFtheory}. Our main focus will 
be on inferring the couplings \eqref{cov-derv-Ftheory_noncl}, \eqref{fieldstrength_braneFtheory}, 
and \eqref{gaugecoupling_F}, which dictate the presence of a non-Abelian gauge symmetry.

The M-theory reduction is performed by using eleven-dimensional supergravity. This implies 
that we have to work with a resolved fourfold $\hat Y_4$. Furthermore, all linearly charged matter
states corresponding to M2-branes on the resolution cycles are integrated out and will not appear 
in the following three-dimensional effective action. 
The starting action is the bosonic part of eleven-dimensional supergravity given by  
\ba
S^{(11)} & = \fr{1}{2} \int \Big( 
\hat  R\,  \hat * 1 - \fr12 \hat G \we \hat * \hat G - \fr16 \hat C \we \hat G \we \hat G \Big) \ ,
\label{11dAction}
\ea
where $\hat R$ is the eleven-dimensional Ricci scalar and $\hat G = d \hat C$ is the four-form 
field strength for the three-form $\hat C$. In the following a hat will indicate that the quantity 
is defined in eleven dimensions.

Clearly, the M-theory reduction 
should not only include harmonic forms, but also contain 
non-closed and exact forms that account for possible gaugings.
These forms can be viewed as parameterising torsion cohomology.  
We thus introduce the two-forms $\o_\S$, and three-forms $(\a_I,\b^I)$ on $\hat Y_4$ that 
need not be harmonic but should be definite eigenstates of the 
Laplace-Beltrami operator. They are related by the non-closure of $\o_\S$ given by 
\ba
d \o_\S & = \til k_{\S}^I \a_I + k_{\S I} \b^I .
\label{omNonClosure} 
\ea
This expression is a generalisation of \eqref{domega_base} in which $\omega_\alpha$ and $\beta^\kappa$
are elements of the base $\cB_3$ of $\hat Y_4$. It also contains the case that $\omega_\Sigma$ yields a gauge field 
of a seven-brane and the non-closure yields the gaugings induced from the 
geometric St\"uckelberg term \eqref{csd7} and \eqref{01Stueck}. For D7-branes this has already been suggested in 
\cite{Grimm:2010ez,Grimm:2011tb,Braun:2014nva}.

Next we introduce the modes of the effective theory that arise from expanding the 
eleven-dimensional metric and the M-theory three-form into $\o_\S$ and $(\a_I$, $\b^I)$. 
We will therefore make an ansatz for the reduction where
\bea
d \hat s^2 & =& g_{\m \n} dx^\m dx^\n + 2 (g_{m \bar n}^0 + i \delta v^\S \o_{\S m \bar n}  ) dy^m dy^{\bar n}\, ,  \\[.1cm]
\hat C & =&  A^\S \wedge\o_\S + \til \x^I \a_I +  \x_I \b^I   \ . \nn
\label{3dAnsatz}
\eea
In this expression $\delta v^\S$, $\til \x^I$ and $\x_I$ are three-dimensional scalar fields, while $A^\S$ are 
three-dimensional vector fields. Note that $\delta v^\S$ parameterise the deformations 
of the Calabi-Yau metric $g_{m \bar n}^0$ that are in general non-K\"ahler. Setting $J = J_0 + \delta v^\S \omega_\S$ one 
has 
\beq
    d J = \delta v^\S d \omega_\S =\delta v^\S \til k_{\S}^I \a_I +  \delta v^\S k_{\S I} \b^I \ ,
\eeq
which implies that there will be a potential induced for the scalars $ \delta v^\S$. 
We will denote the complete three-dimensional scalar potential by $V$, but will refrain discussing 
its precise form. We will also introduce the scalars $v^\S$, which parameterise the expansion 
of $J = v^\S \omega_\S$.  
More important in the following is the reduction of the M-theory three-form part of the 
action. Using \eqref{3dAnsatz} and \eqref{omNonClosure} we see that $\hat G$ is given by 
\ba \label{Ghat}
\hat G & = d A^\S \we \o_\S + D\til\x^I\we \a_I +    D\x_I \we \b^I  + \til \x^I d \a_I + \x_I d \b^I\ .
\ea 
Here we have defined the covariant derivatives
\ba
D\til\x^I &= d \til \x^I -  A^\S \til k_{\S}^I,  & 
D\x_I &= d \x_I -  A^\S k_{\S I }\ .
\ea
As we will show in the following it will be these simple gaugings that are responsible for the gauge structure encountered 
in the F-theory effective action of section \ref{nonA_inFtheory}.

Substituting the ansatz \eqref{3dAnsatz} and \eqref{Ghat} into the action \eqref{11dAction} and 
performing a Weyl rescaling, which puts the effective action in Einstein frame, 
we find the three-dimensional effective theory given by 
\ba \label{3DMreduced_action}
S^{(3)}  =& \fr{1}{2} \int \Big[
R  * 1  - \fr12 G_{\S \L} d L^\S \we * d L^\L - \fr12 G_{\S \L} F^\S \we * F^\L  \nn \\
& - \fr12 \til G_{I J} D \til \x^I \we * D \til \x^J - \fr12  G^{IJ} D  \x_I \we * D  \x_J  
- H_I{}^J   D \til \x^I \we * D \x_J  \nn \\ 
& + \frac{1}{3}M_{\S I}{}^J F^\S \we( \til \x^I D  \x_J-\x_J D \til \x^I ) +\frac{1}{3} M_{\S IJ} F^\S \we  \til\x^I D \til \x^J+\frac{1}{3} M_{\S}{}^{IJ} F^\S \we  \x_I D  \x_J  \nn\\
&  +\frac{1}{3} M_{\S I}{}^J A^\S  \we D \til \x^I\we D \x_J  +\frac{1}{6} M_{\S IJ} A^\S  \we D \til \x^I\we D \til\x^J  +\frac{1}{6} M_{\S}{}^{IJ} A^\S  \we D \x_I\we D \x_J \nn\\
& -\frac{1}{3}N_{\S\L I}\,\til \x^I A^\S\we F^\L +\frac{1}{3}\til N_{\S\L}{}^I\, \x_I A^\S\we F^\L + V*1 \Big]\ .
\ea
The first line contains the kinetic terms for the scalars $v^\S$ and vectors $A^\S$. To write them in this 
simple form, we have used the definitions
\ba
G_{\S \L} & =\cV \int_{\hat Y_4} \o_\S \we * \o_\L \ , & 
L^\S &= \fr{v^\S}{\cV} \ ,
\ea
where $\cV$ is the volume of the manifold $\hat Y_4$. 
To display the couplings of the scalars $(\x_I, \tilde \x^J)$
we have introduced the definitions
\ba
\til G_{I J} & = \fr{1}{\cV} \int_{\hat Y_4} \a_I \we * \a_J \ , & G^{I J} & = \fr{1}{\cV} \int_{\hat Y_4} \b^I \we * \b^J\ , \nn \\
H_{I}{}^{ J} & = \fr{1}{\cV} \int_{\hat Y_4} \a_I \we * \b^J \ , & M_{\S I}{}^J  & =  \int_{\hat Y_4} \o_\S \we \a_I \we \b^J\ , \nn\\
N_{\S\L I} &=\int_{\hat Y_4} \o_\S\we \o_\L\we d\a_I \ , & \til N_{\S\L }^{I} &=-\int_{\hat Y_4}\o_\S\we \o_\L\we d\b^I\ .
\ea
The tensors $N_{\S\L I}$ and $\til N_{\S\L }^{I}$ can be written in terms of the other couplings by integrating by parts and using (\ref{omNonClosure}), which gives
\bea \label{Nexpressions}
N_{\S\L I} &=&k_{\S J}M_{\L I}{}^J+\til k_\S^J M_{\L IJ}+k_{\L J}M_{\S I}{}^J+\til k_\L^J M_{\S IJ}\ , \\
\til N_{\S\L }^{I} &=& k_{\S J}M_{\L }{}^{JI}+\til k_{\S}^JM_{\L J}{}^I+ k_{\L J}M_{\S }{}^{JI}+\til k_{\L}{}^JM_{\S J}{}^I \ , \nn \\\label{kNkN}
k_{\L I}N_{\S\D}^I&=&\til k_{\L}^I N_{\S\D I}\ . \nn
\eea

%It is useful to consider the dual forms of $\o_\S,\,\a_I$ and $\b^I$ which we denote by $\tilde \o^\S,\tilde \b^I$ and $\tilde \a_I$ and satisfy
%
%\ba
%\int \omega_\Sigma\wedge \tilde \omega^\Lambda=\delta_\Sigma^\Lambda, \qquad \int \a_I\wedge \tilde\b^J = \delta_I^J, \qquad \int \tilde \a_I\wedge \b^J= \delta_I^J,
%\ea
%
%together with
%
%\ba
%d\tilde \a_I=k_{\Sigma I}\,\tilde \omega^\Sigma,  &&  d\tilde \b^I=-\tilde k_{\Sigma }^I\,\tilde \omega^\Sigma.
%\ea
%

Let us close this subsection with a few crucial observations. It is straightforward to see that the action \eqref{3DMreduced_action} enjoys an \emph{Abelian} gauge symmetry given by
\ba
\delta A^\S & =d\lam^\S, & 
\delta \til\x^I & = \til k_\S^I \lam^\S, &
 \delta \x_I & = k_{\S I} \lam^\S\ ,
\ea
where $\lam^\S$ is a gauge parameter. However, in the last sections we argued that this system should posses non-Abelian symmetries. Surprisingly, such a non-Abelian structure is present in this setup although it is not obviously realised in terms of the fields we introduced. In the next section we will see how by performing a change of duality frame for certain fields we unravel the non-Abelian symmetries. This new frame turns out to be the correct one in which to perform the F-theory limit and so compare with the four-dimensional effective theory. 

A final comment concerns the supersymmetry properties of the action \eqref{3DMreduced_action}. We
have not demonstrated that this action is indeed supersymmetric. In order to do that one 
would have to introduce complex coordinates on the moduli space and demonstrate 
that the couplings in \eqref{3DMreduced_action} are of special form, e.g.~obtained from a K\"ahler potential. 
This requires the introduction of $(2,1)$-forms on $\hat Y_4$ that can be parameterised by 
a holomorphic function varying over the complex structure moduli space. 
This function is then used in defining the complex coordinates in generalisation of \eqref{def-Psif}
and \eqref{def-Ncoords}. While the ungauged action can then be shown to be supersymmetric, 
it is expected that additional conditions on the allowed gaugings are imposed by supersymmetry.
It would be nice to determine these conditions from a more detailed analysis of the geometry. In the following 
we will continue with our analysis on the bosonic action and focus on the appearance 
of the non-Abelian gaugings manifested through \eqref{cov-derv-Ftheory_noncl} and \eqref{fieldstrengths}.

\subsection{Dualisation of the M-theory effective action} \label{dualizing_action}

The action  \eqref{3DMreduced_action}, obtained by 
dimensional reduction of eleven-dimensional supergravity, is not yet in the duality frame 
that allows a lift to a four-dimensional F-theory configuration. In the following 
we will perform a dualisation to bring it into the correct form.
In order to do this we must first split the three-dimensional fields into those 
which are effected by the duality and those which are not. For this reason we will make 
the decomposition 
\beq
   L^\S  = ( L^\ihat, L^{\a} )\, ,\quad 
   A^\S  = ( A^\ihat, A^{\a} )\, , \quad 
   \til \x^I = ( a^a, \til \x^\k )\, ,  \quad 
   \x_I = ( -b_a,  \x_\k )\ .
\eeq
This is in complete analogy to the ungauged case \cite{Grimm:2010ks}. The multiplet $(L^\a , A^\a)$ will lift to 
the bosonic part of a four-dimensional chiral multiplet with scalars $T_\alpha$ and therefore $A^\a$ needs 
to be dualised into a scalar $\rho_\alpha$ in three dimensions. In contrast $(\til \x^\k,\x_\k)$ will 
comprise the degrees of freedom of a vector in a four-dimensional vector multiplet. These are the 
four-dimensional R-R vectors $A^\kappa$ used in \eqref{cov-derv-Ftheory_noncl}. Therefore one must dualize 
the scalar $\x_\k$ into a three-dimensional vector $A^\k$ before performing the uplift. 
Note that in this section we slightly abuse notation and assert that $A^\k$ and $A^\ihat$ are
three-dimensional vectors. Finally the multiplet $(a^a,b_a)$ lifts to chiral multiplet in the four-dimensional theory which originates from either Type IIB bulk fields, decomposed with respect to internal space two-forms, or from Wilson lines. 

In order to make contact with section \ref{nonA_inFtheory} and 
to keep the discussion simple, we restrict to the case in which 
\ba \label{domega_specialM}
d\omega_\a=k_{\a \kappa}\beta^\kappa,\qquad  \quad d\omega_{\ihat}=\til k_{\ihat}^a\a_a+k_{\ihat a}\beta^a.
\ea
The first condition is the non-closure of forms $\omega_\a$ stemming from the base $\cB_3$
and agrees with \eqref{domega_base}. The second condition accounts for the geometric St\"uckelberg gaugings 
with the seven-brane gauge fields. 
It is important to stress that the dualisation we are preforming only 
works if we impose additional conditions relating the constant 
couplings and gaugings.
Concretely, we find that the duality can be performed when imposing
\bea\label{relations}
\til k_{\ihat }^a M_{\jhat a}{}^\k+k_{\jhat a}M_{\ihat }{}^{a\k} &=&0\, , \nn \\
k_{\ihat  b}M_{\a a}{}^b+k_{\a\k}M_{\ihat  a}{}^\k&=&0\, ,\qquad \tilde k_{\ihat }^b M_{\a ab}=0 \, , \nn \\
\til k_{\ihat }^bM_{\a b}{}^a+k_{\a\k}M_{\ihat }{}^{\k a}&=&0\, ,\qquad k_{\ihat  b}M_\a{}^{ba}=0\, . 
\eea
It is not clear whether these are the weakest conditions that have to be imposed. In particular, it appears 
that imposing only the sum of the expressions in the last two lines, yielding $N_{\ihat  \alpha a} = \tilde N^a_{\ihat  \a} = 0$ by using \eqref{Nexpressions} and \eqref{domega_specialM}, is also sufficient. It would be interesting to 
give a precise geometric reasoning why in an elliptically fibered geometry these vanishing conditions 
are imposed. It appears that these conditions are crucial to perform the F-theory up-lift. This can be 
compared with the vanishing conditions of \cite{Grimm:2011sk,Grimm:2011fx,Cvetic:2013uta} on the $G_4$-flux intersections $\Theta_{\L \S} = \int \omega_\L \wedge \o_\S \wedge G_4$ that need to be imposed for a four-dimensional gauge-invariant setting to exist. 

In addition there is a set of constraints that is readily inferred for an elliptically fibered space by counting the number of legs in the fiber. These are given by
\ba
\label{Mvanishing}
M_{\a \k}{}^\l = M_{\a \k}{}^a = M_{\a a}{}^\k = M_{\a}{}^{\k \l} = M_\a{}^{\k a}  = M_{\a \k \l} = M_{\a \k a} = M_{\ihat}{}^{ \k \l} = M_{\ihat \k \l} = 0 \, ,
\ea
which we will see is true for the duality splitting assignments appropriate for the F-theory lift. 

In order to perform the duality, we proceed in the usual way. We propose a parent Lagrangian that is a function of both the original and dual fields such that it gives back the starting action (\ref{3DMreduced_action}) when we remove the dual fields by using their equations of motion. Alternatively, we can use the equations of motion for the original fields to remove these in favour of the dual ones which gives the dual action. This is a rather complicated computation so we simply quote the result here and refer the reader to appendix \ref{DualAction} for the details. A more detailed analysis 
of this Abelian to non-Abelian duality in various dimension will appear in an upcoming paper \cite{toAppearDuality}. 
The dual Lagrangian is given by
\ba\label{Lagdual}
S  =\,\,& \fr{1}{2} \int \bls  R  * 1  - \fr12 G_{\ihat \jhat} d L^\ihat \we * d L^\jhat- \fr12 G_{\a \b} d L^\a \we * d L^\b- G_{\ihat \a} d L^\ihat \we * d L^\a \\\nn
& - \fr{1}{2}\til G{}_{a}{}_{b} Da{}^{a}\w*Da{}^{b}
   - \fr{1}{2}\til G{}^{a}{}^{b} Db{}_{a}\w*Db{}_{b} 
    +\til H{}_{a}{}^{b}Da{}^{a}\w*Db{}_{b}  
    - \fr{1}{2}G^{-1}{}{}^{\a}{}^{\b}\hat D \r{}_{\a}\w*\hat D \r{}_{\b}  \\\nn
& - \fr{1}{2}\til G{}_{\k}{}_{\l} D\til\x{}^{\k}\w*D\til\x{}^{\l} 
    -  \til G{}_{a}{}_{\k}Da{}^{a}\w*D\til\x{}^{\k} 
     +\til H{}_{\k}{}^{a}Db{}_{a}\w*D\til\x{}^{\k} 
    - \fr{1}{2}G^{-1}{}{}_{\k}{}_{\l}U{}^{\k}\w*U{}^{\l}  \\\nn
&  - \fr{1}{2}\til G{}_{\ihat }{}_{\jhat }F{}^{\ihat }\w*F{}^{\jhat } 
    - \fr{1}{3}M{}_{\ihat }{}_{a}{}_{b}a{}^{a}Da{}^{b}\w F{}^{\ihat }
    - G^{-1}{}{}_{\k}{}_{\l}G{}^{a}{}^{\k}Db{}_{a}\w U{}^{\l} 
    + G^{-1}{}{}_{\k}{}_{\l}H{}_{a}{}^{\k}Da{}^{a}{}\w U{}^{\l}\\\nn
&   +G^{-1}{}{}_{\l}{}_{\h}H{}_{\k}{}^{\l} D\til\x{}^{\k}\w U{}^{\h} - G^{-1}{}{}^{\a}{}^{\b}G{}_{\b}{}_{\ihat }\hat D \r{}_{\a}\w F{}^{\ihat } + \fr{1}{3}M{}_{\ihat }{}_{\k}{}^{a}\til\x{}^{\k}Db{}_{a}\w F{}^{\ihat } \\\nn
& + \fr{1}{3}M{}_{\ihat }{}_{a}{}^{b}a{}^{a}Db{}_{b}\w F{}^{\ihat } 
- \fr{1}{3}M{}_{\ihat }{}_{a}{}_{\k}a{}^{a}D\til\x{}^{\k}\w F{}^{\ihat } 
- \fr{1}{3}M{}_{\ihat }{}_{b}{}^{a}b{}_{a}Da{}^{b}\w F{}^{\ihat } 
 \\\nn
&   - \fr{1}{3}M{}_{\ihat }{}^{a}{}^{b}b{}_{a}Db{}_{b}\w F{}^{\ihat }
     - \fr{1}{3}M{}_{\ihat }{}_{\k}{}^{a}b{}_{a}D\til\x{}^{\k}\w F{}^{\ihat } 
     + \fr{1}{3}M{}_{\ihat }{}_{a}{}_{\k}\til\x{}^{\k}Da{}^{a}\w F{}^{\ihat }\\\nn
&- \fr{1}{3}M{}_{\ihat }{}_{a}{}_{\k}A{}^{\ihat }\w Da{}^{a}\w D\til\x{}^{\k} - \fr{1}{3}M{}_{\ihat }{}_{\k}{}^{a}A{}^{\ihat }\w Db{}_{a}\w D\til\x{}^{\k} + \fr{1}{3}M{}_{\ihat }{}_{a}{}^{b}A{}^{\ihat }\w Da{}^{a}\w Db{}_{b}\\\nn
& - \fr{1}{6}M{}_{\ihat }{}^{a}{}^{b}A{}^{\ihat }\w Db{}_{a}\w Db{}_{b} 
  - \fr{1}{6}M{}_{\ihat }{}_{a}{}_{b}A{}^{\ihat }\w Da{}^{a}\w Da{}^{b}  + V*1 \brs \, , 
\ea
where, as promised, we traded the scalars $\x_\k$ and vectors $A^\a$, for the vectors $A^\k$ and scalars $\r_\a$ respectively. It should be stressed that for simplicity we did not take into account the moduli 
dependence of the coupling functions. In particular, we have frozen the complex structure moduli in 
all these considerations. 
The coupling functions appearing in the dualised action \eqref{Lagdual} are defined as
\be
\begin{array}{ll}
\til H{}_{a}{}^{b}=H{}_{a}{}^{b} -G{}^{b}{}^{\l}G^{-1}{}{}_{\k}{}_{\l}H{}_{a}{}^{\k},\qquad  &\qquad  \til G{}_{a}{}_{\k}=\til G{}_{a}{}_{\k} -G^{-1}{}{}_{\l}{}_{\h}H{}_{a}{}^{\l}H{}_{\k}{}^{\h} , \\
\til H{}_{\k}{}^{a}=H{}_{\k}{}^{a}-G{}^{a}{}^{\h}G^{-1}{}{}_{\l}{}_{\h}H{}_{\k}{}^{\l},\qquad  &\qquad \til G{}_{a}{}_{b}=\til G{}_{a}{}_{b} - G^{-1}{}{}_{\k}{}_{\l}H{}_{a}{}^{\k}H{}_{b}{}^{\l} , \\
\til G{}^{a}{}^{b} =G{}^{a}{}^{b} - G{}^{a}{}^{\k}G{}^{b}{}^{\l}G^{-1}{}{}_{\k}{}_{\l},\qquad  &\qquad \til G{}_{\k}{}_{\l}=\til G{}_{\k}{}_{\l} - G^{-1}{}{}_{\h}{}_{\r}H{}_{\k}{}^{\h}H{}_{\l}{}^{\r} , \\
\til G{}_{\ihat }{}_{\jhat } = G{}_{\ihat }{}_{\jhat }-G^{-1}{}{}^{\a}{}^{\b}G{}_{\a}{}_{\ihat }G{}_{\b}{}_{\jhat }\, .
\end{array}
\ee
The action includes the \textit{gauge-invariant} expressions
\bea\label{hat}
\hat D\rho_\a&=&D\rho_\a+\frac{1}{2}M_{\a a}{}^b(a^aDb_b-b_bDa^a)-\frac{1}{2}M_{\a ab}a^aDa^b-\frac{1}{2}M_{\a}{}^{ab}b_aDb_b\,, \\
U^\k&=&F^\k+M_{\ihat \lam}{}^\k\tilde\xi^\lam F^{\ihat }+M_{\ihat a}{}^\k a^a F^{\ihat }-M_{\ihat }{}^{a\k} b_a F^{\ihat }\, , 
\eea
where
\bea\label{covrho}
D\rho_\a&=&d\rho_\a-k_{\a\k}A^\k-\frac{1}{2}k_{\a\k}M_{\ihat a}{}^\k a^a A^{\ihat }+\frac{1}{2}k_{\a\k}M_{\ihat }{}^{a\k}b_a A^{\ihat }\, , \\
F^\k&=&dA^\k+\frac{1}{2}(\tilde k_{\jhat }^aM_{\ihat a}{}^\k+ k_{\jhat a}M_{\ihat }{}^{a\k})A^{\ihat }\wedge A^{\jhat }.
\eea
We expect that the non-Abelian structure is linked to the coupling $M_{\a a}{}^b$. This is not obvious from the expressions above but we may use the relations (\ref{relations}) to make it manifest, namely
\bea\nn
D\rho_\a&=&d\rho_\a-k_{\a\k}A^\k+\frac{1}{2}M_{\a a}{}^b (k_{\ihat b}a^a-\til k_{\ihat }^a b_b)A^{\ihat }\, ,\\
k_{\a\k}F^\k&=&k_{\a\k}dA^\k+\tilde k_{\jhat }^a k_{\ihat b}M_{\a a}{}^bA^{\ihat }\wedge A^{\jhat } \, . 
\eea
This also shows that the field strength satisfies the projection condition \eqref{fieldstrengths} that was required for closure of the gauged subalgebra.

Let us now split the index $\ihat $ further into $(0,i)$. This allows us to denote $\o_0$ as the two-form Poincar\'e dual to the base of the elliptic fibration (which we assume to be closed in the following), $\o_i$ as dual to blow-up divisors and $\o_\a$ as dual to vertical divisors. Similarly we now understand the splitting $\a_I = (\a_a, \a_\k)$ and $\b^I = (\b^a, \b^\k)$ as being such that $\a^a$ and $\b_a$ have a component with one leg in the fiber while $\a_\k$ and $\b^\k$ have legs only in the base directions. 
This decomposition justifies the constraints \eqref{Mvanishing} which may be seen by counting legs of the forms present. We also now impose that 
\ba
M_{0\k}{}^\l&=\delta_\k^\l \, , &
M_{i \k}{}^\l&= 0  \, , 
\ea
the first of which shows that $\a_\k$ and $\b^\k$ form a symplectic basis for three-forms on the base. 
With this decomposition we see that the gaugings decompose as $\til k_\ihat^a = ( 0 , \til k_i^a)$ and  $ k_{\ihat a} = ( 0 ,  k_{i a})$.

Having performed this further decomposition the field strengths and covariant derivatives may be written as
\ba
\label{CovDsandFieldStrenghtsExpanded}
U^\k&=F^\k+(\tilde\xi^\k +M_{0a}{}^\k a^a-M_{0}{}^{a\k} b_a)F^{0}+M_{ ia}{}^\k a^a F^{i}-M_{ i}{}^{a\k} b_a F^{ i} \,, \nn \\
F^\k&= dA^\k+\frac{1}{2}(\tilde k_{j }^aM_{i a}{}^\k+ k_{j a}M_{i }{}^{a\k})A^{i}\wedge A^{j }\, ,  \nn \\
\hat D\rho_\a&= D\rho_\a+\frac{1}{2}M_{\a a}{}^b(a^aDb_b-b_bDa^a)-\frac{1}{2}M_{\a ab}a^aDa^b-\frac{1}{2}M_{\a}{}^{ab}b_aDb_b \, , \nn \\
D\rho_\a&= d\rho_\a-k_{\a\k}A^\k+\frac{1}{2}M_{\a a}{}^b (k_{ ib}a^a-\til k_{ i}^a b_b)A^{ i}\, . 
\ea
From these expressions we clearly see that a non-Abelian gauge symmetry has emerged after the dualisation. In particular, only the field strength $F^\k$ includes the usual non-Abelian term $A^i \wedge A^j$ so that together $\{F^i, F^\k\}$ correspond to the field strengths of the extended Heisenberg algebra.

To close this section let us discuss the gauge coupling function in some detail. 
As already mentioned, the Heisenberg group is both non-compact and non-semisimple so the kinetic terms for the gauge bosons cannot be proportional to the Killing form. From \eqref{Lagdual} and using the definition for $U^\k$ in \eqref{CovDsandFieldStrenghtsExpanded} we can read off these kinetic terms as %
\bea\nn
G^{-1}{}{}_{\k}{}_{\l} U^\k\w*U^\l+\til G_{ i j} F^{ i}\w*F^{ j}=  G^{-1}{}{}_{\k}{}_{\l} F^\k\w*F^\l+2G^{-1}{}{}_{\k}{}_{\l}(M_{ ia}{}^\k a^a-M_{ i}{}^{a\k} b_a)F^\l\w*F^{ i}\\\nn
(G^{-1}{}{}_{\k}{}_{\l}(M_{ ia}{}^\k a^a-M_{ i}{}^{a\k} b_a)(M_{ ja}{}^\l a^a-M_{ j}{}^{a\l} b_a)+\til G_{ i j} )F^{ i}\w*F^{ j} \, ,
\eea
where here we have set $A^0$ to zero in order to focus only on a particular set of terms. The gauge kinetic function that we see here is independent of the gaugings that we have introduced so corresponds to the supersymmetric result that is also present in the Calabi-Yau fourfold reduction. We also note that it contains the scalars $a^a$ and $b_a$ in a way that causes it to transform under the gauge symmetries. It is then clear that the constraints \eqref{relations} are needed in order to ensure that the transformation of $a^a$ and $b_a$ in the gauge kinetic function cancels the variation of $F^i$ and $F^\k$ and so leaves these terms invariant. When the F-theory limit is taken and these kinetic terms are lifted to the corresponding four-dimensional effective theory, this property must be preserved. In addition to this the the gauge kinetic function must become a holomorphic function of the complex coordinates, in order for the action to be supersymmetric.

\section{Conclusions}

In this paper we discussed the appearance of discrete non-Abelian gauge symmetries in Type IIB compactifications 
to four space-time dimensions. We first reviewed the relationship between discrete symmetries and the gaugings of the isometries of the scalar manifold. We then analysed the symmetries in weakly coupled Type IIB orientifold 
compactifications that are captured by a generalisation of the Heisenberg algebra. We  turned 
to the gaugings and argued that, when including a D7-brane sector, it appears impossible 
to find non-Abelian discrete symmetries at weak string coupling. In orientifold reductions with torsion homology 
we argued that non-Abelian discrete symmetries appear to be in tension with simple 
supersymmetry considerations. Having carried this out we suggested a concrete scenario which demonstrated that non-Abelian discrete symmetries can 
arise in more general F-theory compactifications with mutually non-local seven-branes. Remarkably, these considerations 
 require the use of the full power of F-theory away from the weak coupling limit. We argued that 
the gauge fields on general $(p,q)$-seven-branes can gauge both R-R and NS-NS axions yielding a non-Abelian 
gauge structure generalising the Heisenberg algebra. The sources of these gaugings where identified to be: (1) geometric St\"uckelberg terms on $(p,q)$-seven-branes, (2) fluxes on seven-branes, (3) torsion three-form cohomology in the six-dimensional compactification space $\cB_3$. It was a non-trivial task to confirm these statements using the duality between M-theory and F-theory. Importantly this required the dualisation of an Abelian theory into a non-Abelian theory in three space-time dimensions. 

We have argued that there is a setting in which all fields associated with the gaugings we described arise from seven-branes. 
To make this more precise one can follow the strategy of \cite{Anderson:2014yva,Garcia-Etxebarria:2014qua}. In these works it 
was suggested that for Abelian groups the degrees of freedom in the non-linearly 
charged $N^a$ can be captured by open string degrees of freedom $\phi$ linearly 
charged under the Abelian group. It is natural to conjecture that one can proceed analogously 
for the non-Abelian configurations considered in this work. For the Heisenberg group such linearly
charged states $\phi$ are, for example, given by the theta representation. However, note that these representations of 
the continuous Heisenberg group are infinite dimensional. At first, this appears to be at odds with the interpretation of
$\phi$ as a matter state on intersecting seven-branes. 
However, the theta representations of the \textit{discrete} non-Abelian group can 
be finite dimensional. Recalling that we have found that there is no vacuum of our theory in which the continuous symmetry 
is unbroken it might therefore be the case that geometrically only the discrete non-Abelian group is realised. 
Our analysis suggests that it might be possible to find geometries with intersecting seven-branes
that have matter linearly charged under a discrete Heisenberg group \cite{BerasaluceGonzalez:2012vb,Marchesano:2013ega}.  
The non-Abelian nature of the gaugings then might be tied to the requirement that string junctions between certain seven-branes, as for example (1,0)- and (0,1)-branes, have to exist and end on a third seven-brane. 
We leave a deeper investigation of such seven-brane settings to future work.

It is interesting to summarise the complications that we had to face in our analysis. Firstly, one could have thought that a straightforward generalisation of the reductions with torsion homology  \cite{BerasaluceGonzalez:2012vb} leads to Calabi-Yau fourfold reductions with the desired non-Abelian structure. However, an explicit computation shows that this is not the case. More precisely, a direct reduction of eleven-dimensional supergravity formulated with the three-form field yields only Abelian gaugings even when including torsional cohomology. The non-trivial observation is, however, that this direct reduction is not 
yet in the correct duality frame to perform the lift to F-theory. After performing the duality, non-Abelian gaugings arise 
and allow us to identify  genuine F-theory gaugings in settings with $(p,q)$-seven-branes. 
Secondly, showing consistency with supersymmetry turned out to be a non-trivial task which we will to return to. Indeed, in the Type IIB analysis of section \ref{susygaugings_ori} we found that the reduction considered is not supersymmetric. The local form of the $\cN=1$
complex moduli space arising in a general F-theory setting dictates constraints on the allowed holomorphic gaugings. 

  Let us close by highlighting the intriguing observation we made concerning the gauge coupling 
 functions in the considered F-theory effective actions. 
 If one is able to gauge a non-compact and non-semisimple non-Abelian group, such as
 the extensions of the Heisenberg groups we found in our settings, then one necessarily has to have 
 a gauge coupling function depending on the complex scalar fields that are charged. In fact, this 
 dependence will by partly dictated by the gauge invariance of the action. 
 We have shown that this consistency requirement is automatically satisfied for 
 the F-theory settings we considered. Interestingly, in our setups the modifications of the gauge coupling 
 function are independent of the parameters determining the subgroup of the isometry group 
 that is gauged.  One can thus infer properties of the gauge coupling functions
 in this F-theory reduction by analysing the isometries of the scalar manifold. 
 We have checked that the required modifications give precisely the kinetic mixing terms in 
 standard Type IIB reduction with D7-branes. It would be interesting to understand 
 if this link between holomorphic isometries and the form of the gauge coupling function 
 is a general feature of string theory effective actions.

\acknowledgments

We are  grateful to Sebastian Greiner, Olaf Hohm, Jan Keitel, Fernando Marchesano and especially I\~naki Garc\'ia-Etxebarria for illuminating discussions. This work was supported by a grant from the Max Planck Society.

\appendix

\section{Dualisation of the three-dimensional action}
\label{DualAction}

We wish to perform the dualisation of the action  \eqref{3DMreduced_action} and to put the fields of the action into a frame that is appropriate for the F-theory lift. In order to simplify the analysis we will first freeze out the moduli dependence of $G_{\S \L}$, $\til G_{I J}$, $G^{IJ}$ and $H_I{}^J$. We will also make use of \eqref{Nexpressions} to remove $N_{\S \L}{}^{I}$ and $N_{\S \L I }$ from the action. We will then split the index on each field such that those that are to be dualised are identified from those which are not.  This will be carried out by splitting
 \beq
   L^\S  = ( L^\ihat, L^{\a} )\, ,\quad 
   A^\S  = ( A^\ihat, A^{\a} )\, , \quad 
   \til \x^I = ( a^a, \til \x^\k )\, ,  \quad 
   \x_I = ( -b_a,  \x_\k )\ .
\eeq
where the fields $\x_\k$ and $A^\a$ are to be dualised. With this splitting we will also restrict the gaugings as shown in  \eqref{domega_specialM} and \eqref{relations} so that the covariant derivatives and field strengths are given by 
\ba
D a^a & = d a^a - \til k^a_{\ihat} A^\ihat \, , &  
D b_a & = d b_a + k_{\ihat a } A^\ihat \, , \nn \\ 
D \til \x^\k & = d \x^\k \, , &  
D \x_\k & = d \x_\k - k_{\a \k}  A^\a \, , \nn \\
F^\ihat & = d A^\ihat \, , & 
F^\a & = d A^\a \, ,
\ea
and will restrict $M_{\S I}{}^J$ as shown in \eqref{Mvanishing}.
Performing these steps gives the starting action on which we will perform the duality, given by 
\ba\label{Lag1}
S  =\,\,& \fr{1}{2} \int \bls  R  * 1  - \fr12 G_{\ihat \jhat} d L^\ihat \we * d L^\jhat- \fr12 G_{\a \b} d L^\a \we * d L^\b- G_{\ihat \a} d L^\ihat \we * d L^\a \\\nn
& + H{}_{a}{}^{b}Da{}^{a}{}\w*Db{}_{b}-\til G{}_{a}{}_{\k}  Da{}^{a}\w*D\til\x{}^{\k}- H{}_{a}{}^{\k}Da{}^{a}\w*D\x{}_{\k} +H{}_{\k}{}^{a} Db{}_{a}\w*D\til\x{}^{\k}  \\\nn
&- H{}_{\l}{}^{\k} D\x{}_{\k}\w*D\til\x{}^{\l}- \fr{1}{2}\til G{}_{a}{}_{b}Da{}^{a}\w*Da{}^{b} - \fr{1}{2}G{}^{a}{}^{b}Db{}_{a}\w*Db{}_{b} - \fr{1}{2}\til G{}_{\k}{}_{\l}D\til\x{}^{\k}\w*D\til\x{}^{\l} \\\nn
&- \fr{1}{2}G{}^{\k}{}^{\l} D\x{}_{\k}\w*D\x{}_{\l}- G{}_{\a}{}_{{\ihat}}F{}^{\a}\w*F{}^{{\ihat}} + \fr{1}{3}M{}_{\a}{}_{a}{}^{b}A{}^{\a}\w Da{}^{a}\w Db{}_{b}+ G{}^{a}{}^{\k}Db{}_{a}\w*D\x{}_{\k}  \\\nn
&+ \fr{1}{3}M{}_{\ihat}{}_{a}{}^{b}A{}^{\ihat}\w Da{}^{a}\w Db{}_{b}- \fr{1}{3}M{}_{\ihat}{}_{a}{}_{\k} A{}^{\ihat}\w Da{}^{a}\w D\til\x{}^{\k}- \fr{1}{3}M{}_{\ihat}{}_{a}{}^{\k}A{}^{\ihat}\w Da{}^{a}\w D\x{}_{\k}\\\nn
& - \fr{1}{3}M{}_{\ihat}{}_{\k}{}^{a} A{}^{\ihat}\w Db{}_{a}\w D\til\x{}^{\k}+ \fr{1}{3}M{}_{\ihat}{}^{a}{}^{\k}A{}^{\ihat}\w Db{}_{a}\w D\x{}_{\k} + \fr{1}{3}M{}_{\ihat}{}_{\l}{}^{\k}A{}^{\ihat}\w D\x{}_{\k}\w D\til\x{}^{\l} \\\nn
&- \fr{1}{2}G{}_{\a}{}_{\b}F{}^{\a}\w*F{}^{\b} - \fr{1}{2}G{}_{\ihat}{}_{\jhat}F{}^{\ihat}\w*F{}^{\jhat} - \fr{1}{3}M{}_{\a}{}_{a}{}_{b}a{}^{a}Da{}^{b}\w F{}^{\a}- \fr{1}{3}M{}_{\ihat}{}_{a}{}_{b}a{}^{a}Da{}^{b}\w F{}^{\ihat} \\\nn
& + \fr{1}{3}M{}_{\a}{}_{a}{}^{b}a{}^{a}Db{}_{b}\w F{}^{\a} + \fr{1}{3}M{}_{\ihat}{}_{a}{}^{b}a{}^{a}Db{}_{b}\w F{}^{\ihat}- \fr{1}{3}M{}_{\ihat}{}_{a}{}_{\k} a{}^{a}D\til\x{}^{\k}\w F{}^{\ihat}- \fr{1}{3}M{}_{\ihat}{}_{a}{}^{\k}a{}^{a}D\x{}_{\k}\w F{}^{\ihat} \\\nn
& - \fr{1}{6}M{}_{\a}{}^{a}{}^{b}A{}^{\a}\w Db{}_{a}\w Db{}_{b} - \fr{1}{6}M{}_{\ihat}{}_{a}{}_{b} A{}^{\ihat}\w Da{}^{a}\w Da{}^{b}- \fr{1}{6}M{}_{\ihat}{}^{a}{}^{b}A{}^{\ihat}\w Db{}_{a}\w Db{}_{b}\\\nn
& - \fr{1}{6}M{}_{\a}{}_{a}{}_{b}A{}^{\a}\w Da{}^{a}\w Da{}^{b}- \fr{1}{3}M{}_{\a}{}_{b}{}^{a} b{}_{a}Da{}^{b}\w F{}^{\a}- \fr{1}{3}M{}_{\ihat}{}_{b}{}^{a}b{}_{a}Da{}^{b}\w F{}^{\ihat} \\\nn
&- \fr{1}{3}M{}_{\ihat}{}^{a}{}^{b} b{}_{a}Db{}_{b}\w F{}^{\ihat}- \fr{1}{3}M{}_{\ihat}{}_{\k}{}^{a}b{}_{a}D\til\x{}^{\k}\w F{}^{\ihat} + \fr{1}{3}M{}_{\ihat}{}^{a}{}^{\k}b{}_{a}D\x{}_{\k}\w F{}^{\ihat}- \fr{1}{3}M{}_{\a}{}^{a}{}^{b}b{}_{a}Db{}_{b}\w F{}^{\a} \\\nn
&+ \fr{1}{3}M{}_{\ihat}{}_{a}{}_{\k}\til\x{}^{\k}Da{}^{a}\w F{}^{\ihat} + \fr{1}{3}M{}_{\ihat}{}_{\k}{}^{a}\til\x{}^{\k}Db{}_{a}\w F{}^{\ihat} + \fr{1}{3}k{}_{\a}{}_{\k}M{}_{\ihat}{}_{\l}{}^{\k}\til\x{}^{\l}A{}^{\a}\w F{}^{\ihat} \\\nn
& + \fr{1}{3}k{}_{\a}{}_{\k}M{}_{\ihat}{}_{\l}{}^{\k}\til\x{}^{\l}A{}^{\ihat}\w F{}^{\a} - \fr{1}{3}M{}_{\ihat}{}_{\l}{}^{\k}\til\x{}^{\l}D\x{}_{\k}\w F{}^{\ihat} + \fr{1}{3}M{}_{\ihat}{}_{a}{}^{\k}\x{}_{\k}Da{}^{a}\w F{}^{\ihat} \\\nn
&- \fr{1}{3}M{}_{\ihat}{}^{a}{}^{\k}\x{}_{\k}Db{}_{a}\w F{}^{\ihat}+ \fr{1}{3}M{}_{\ihat}{}_{\k}{}^{\l}\x{}_{\l}D\til\x{}^{\k}\w F{}^{\ihat}+ V*1 \brs \, . 
\ea
This action has a purely Abelian set of gauge symmetries. 

Next let us define the projectors $\Pi_{\a}^{\b}$ and $\Pi^{\k}_{\l}$ which allow us to identify the fields that participate in the gaugings. These satisfy
\ba
\Pi_{\a}^{\b}\, k_{\b \k} & = k_{\a \k} \,, & 
 k_{\a \l}\, \Pi^{\l}_{ \k}  & =  k_{\a \k} \, ,  \nn \\
\Pi_{\a}^{\g}\, \Pi_{\g}^{\b} &= \Pi_{\a}^{\b} \, ,  & 
\Pi^{\k}_{ \d} \, \Pi^{\d}_{ \l}  & = \Pi^{\k}_{ \l} \, ,
\label{ProjectorProperties}
\ea
and may be constructed using the so-called Moore-Penrose pseudo-inverse of the matrix $k_{\a\k}$ which we denote by $\inv^{\k\a}$. Then,
\ba
\Pi_{\a}^{\b} &=k_{\a\k}\inv^{\k\b}& \Pi^{\l}_{ \k}&=\inv^{\k\a}k_{\a\l} \, . 
\ea
In addition to these constraints we will also demand that the projectors satisfy certain symmetry conditions such that 
\ba
\Pi_{\a}^{\g} G_{\g \b} & = \Pi_{\b}^{\g} G_{\g \a}   & 
\Pi^{\k}_{ \h} G^{\h \l} & = \Pi^{\l}_{ \h} G^{\h \k}  
\ea
These conditions make the pseudo-inverse $\inv^{\k\a}$ unique for a given $k_{\a\k}$.
For convenience we will also define the projectors in the orthogonal directions  given by 
\ba
\Pperp_{\a}^{\b} & = ( \d_{\a}{}^\b - \Pi_{\a}^{\b} ) \, , & 
\Pperp_{\k}^{\l} & = ( \d_{\k}{}^\l - \Pi_{\k}^{\l} )  \, .
\ea

Having defined these quantities we are now in a position to propose the parent action, from which we will deduce the dual. This is given by,
\ba\label{LagParent}
S  =\,\,& \fr{1}{2} \int \bls  R  * 1  - \fr12 G_{\ihat \jhat} d L^\ihat \we * d L^\jhat- \fr12 G_{\a \b} d L^\a \we * d L^\b- G_{\ihat \a} d L^\ihat \we * d L^\a \\\nn
&+ (G^{-1}{}{}_{\h}{}_{\s}H{}_{a}{}^{\l}H{}_{\k}{}^{\s}H{}_{a}{}^{b}-G{}^{b}{}^{\l}G^{-1}{}{}_{\k}{}_{\h}H{}_{a}{}^{\k}) \Pperp{}_{\l}^{\h}Da{}^{a}\w*Db{}_{b}- H{}_{a}{}^{\l}\Pi{}^{\k}_{\l}Da{}^{a}\w*D\x{}_{\k}\\\nn
& + (G^{-1}{}{}_{\l}{}_{\r}H{}_{a}{}^{\l}H{}_{\k}{}^{\h}\Pperp{}^{\r}_{\h}- \til G{}_{a}{}_{\k})Da{}^{a}\w*D\til\x{}^{\k}\\\nn
&- G{}^{a}{}^{\h}G^{-1}{}{}_{\h}{}_{\s}H{}_{\k}{}^{\l}\Pperp{}_{\l}^{\s}Db{}_{a}\w*D\til\x{}^{\k} + H{}_{\k}{}^{a} Db{}_{a}\w*D\til\x{}^{\k}+ G{}^{a}{}^{\k}Db{}_{a}\w*D\x{}_{\k} \\\nn
&- G{}^{a}{}^{\l}\Pperp{}^{\k}_{\l}Db{}_{a}\w*D\x{}_{\k} + H{}_{\l}{}^{\h}\Pperp{}^{\k}_{\h}D\x{}_{\k}\w*D\til\x{}^{\l} -H{}_{\l}{}^{\k}  D\x{}_{\k}\w*D\til\x{}^{\l}\\\nn
& -G{}_{\b}{}_{\g}\Pperp{}^{\a}{}^{\g} M{}_{\a}{}_{b}{}^{a}b{}_{a}Da{}^{b}\w F{}^{\b} + \fr{1}{2}G^{-1}{}{}_{\k}{}_{\h}H{}_{a}{}^{\k}H{}_{b}{}^{\l}\Pperp{}^{\h}_{\l}Da{}^{a}\w*Da{}^{b}\\\nn
& - \fr{1}{2}\til G{}_{a}{}_{b}Da{}^{a}\w*Da{}^{b} + G^{-1}{}{}_{\h}{}_{\r}H{}_{a}{}^{\k}\Pperp{}_{\k}^{\h}Da{}^{a}\w U{}^{\r} - \fr{1}{2}G{}^{a}{}^{b}Db{}_{a}\w*Db{}_{b}\\\nn
& + \fr{1}{2}Db{}_{a}\w*Db{}_{b}G{}^{a}{}^{\k}G{}^{b}{}^{\l}G^{-1}{}{}_{\k}{}_{\h}\Pperp{}^{\h}_{\l} - Db{}_{a}G{}^{a}{}^{\k}G^{-1}{}{}_{\h}{}_{\r}P_{\k}^{\h}\w U{}^{\r} \\\nn
&+ \fr{1}{2}D\til\x{}^{\k}\w*D\til\x{}^{\l}G^{-1}{}{}_{\h}{}_{\s}H{}_{\k}{}^{\h}H{}_{\l}{}^{\r}\Pperp{}^{\s}_{\r} - \fr{1}{2}D\til\x{}^{\k}\w*D\til\x{}^{\l}\til G{}_{\k}{}_{\l} \\\nn
&+ D\til\x{}^{\k}G^{-1}{}{}_{\l}{}_{\h}H{}_{\k}{}^{\l}\Pperp{}^{\h}_{\s}\w U{}^{\s} - \fr{1}{2}G^{\k\r}\Pi_{\r}^{\l}D\x{}_{\k}\w*D\x{}_{\l} + D\x{}_{\k}\Pperp{}^{\k}_{\h}\w U{}^{\h} - D\x{}_{\k}\Pi{}^{\k}_{\h}\w U{}^{\h} 
 \\ \nn
&- \hat D \til\r{}_{\a}\Pperp{}^{\a}_{\b}\w F{}^{\b} - \hat D \til\r{}_{\a}G{}_{\b}{}_{\ihat}G^{-1}{}^{\a\b}\w F{}^{\ihat}- \fr{1}{2}G^{-1}{}^{\a\b}\hat D \til\r{}_{\a}\w*\hat D \til\r{}_{\b}\\\nn
& +\Pperp{}^{\h}_{\l}M{}_{\ihat}{}_{\k}{}^{\l} \x{}_{\h}D\til\x{}^{\k}\w F{}^{\ihat}-\Pperp{}^{\k}_{\l}M{}_{\ihat}{}^{a}{}^{\l} \x{}_{\k}Db{}_{a}\w F{}^{\ihat}+ \Pperp{}^{\l}_{\k}M{}_{\ihat}{}_{a}{}^{\k}\x{}_{\l}Da{}^{a}\w F{}^{\ihat} \\\nn
&+ \fr{1}{3}M{}_{\ihat}{}_{a}{}^{b}A{}^{\ihat}\w Da{}^{a}\w Db{}_{b} - \fr{1}{3}M{}_{\ihat}{}_{a}{}_{\k}A{}^{\ihat}\w Da{}^{a}\w D\til\x{}^{\k} - \fr{1}{3}M{}_{\ihat}{}_{\k}{}^{a}A{}^{\ihat}\w Db{}_{a}\w D\til\x{}^{\k}\\\nn
& - \fr{1}{2}\Pperp{}_{\a}^{\b}M{}_{\b}{}_{a}{}_{b} a{}^{a}Da{}^{b}\w F{}^{\a}- \fr{1}{2}\Pperp{}_{\a}^{\b}M{}_{\b}{}^{a}{}^{b}b{}_{a}Db{}_{b}\w F{}^{\a}
\ea
\ba \nn
& - \fr{1}{2}\Pperp{}_{\l}^{\h}G^{-1}{}{}_{\h}{}_{\r}U{}^{\l}\w*U{}^{\r} + \fr{1}{2}G{}_{\a}{}_{\ihat}G{}_{\b}{}_{\jhat}G^{-1}{}^{\a\b}F{}^{\ihat}\w*F{}^{\jhat}- \fr{1}{2}G{}_{\ihat}{}_{\jhat}F{}^{\ihat}\w*F{}^{\jhat}- \fr{1}{3}M{}_{\ihat}{}_{a}{}_{b}a{}^{a}Da{}^{b}\w F{}^{\ihat}\\\nn
& + \fr{1}{3}M{}_{\ihat}{}_{a}{}^{b}a{}^{a}Db{}_{b}\w F{}^{\ihat}- \fr{1}{3}M{}_{\ihat}{}_{a}{}_{\k} a{}^{a}D\til\x{}^{\k}\w F{}^{\ihat}- \fr{1}{6}M{}_{\ihat}{}_{a}{}_{b}A{}^{\ihat}\w Da{}^{a}\w Da{}^{b}\\\nn
& - \fr{1}{6}M{}_{\ihat}{}^{a}{}^{b} A{}^{\ihat}\w Db{}_{a}\w Db{}_{b}- \fr{1}{3}M{}_{\ihat}{}_{b}{}^{a} b{}_{a}Da{}^{b}\w F{}^{\ihat}- \fr{1}{3}M{}_{\ihat}{}^{a}{}^{b}b{}_{a}Db{}_{b}\w F{}^{\ihat}\\\nn
& - \fr{1}{3}M{}_{\ihat}{}_{\k}{}^{a} b{}_{a}D\til\x{}^{\k}\w F{}^{\ihat}+ \fr{1}{3}M{}_{\ihat}{}_{a}{}_{\k} \til\x{}^{\k}Da{}^{a}\w F{}^{\ihat}+ \fr{1}{3}M{}_{\ihat}{}_{\k}{}^{a}\til\x{}^{\k}Db{}_{a}\w F{}^{\ihat} \brs \, . 
\ea
In this action the quantities $U^\k$ and $\hat D \rho_\a$ are not a priori field strengths and covariant derivatives but are instead given by
\ba
U^\k & = \Pi_\l^\k d B^\l +  \Pperp_\l^\k H^\l +  \fr12 M_{\ihat a}{}^\k F^\ihat a^a  + \fr12 M_{\ihat a}{}^\k  A^\ihat Da^a 
\nn \\& 
+  \fr12 M_{i }{}^{a \k} F^i b_a  + \fr12 M_{\ihat }{}^{a \k}  A^\ihat Db_a + M_{\ihat \l}{}^\k F^\ihat \til \x^\l \, ,  \nn \\ 
\hat D \rho_\a & = \Pi_\a^\b d \r_\b +  \Pperp_\a^\b h_\b  - \fr12 k_{\a \k} B^\k - \fr12 M_{\b a b} a^a D a^b 
 + \fr12 M_{\a a}{}^{ b} a^a D b_b 
 \nn \\&
 -  \fr12 M_{\a a}{}^{ b} b_b D a^a  -  \fr12 M_{\a}{}^{ a b} b_a D b_b)  \, , 
\ea
where the fundamental variables in \eqref{LagParent} are treated as being the variables of \eqref{Lag1} as well as $B^\k$, $H^\k$, $\r_\b$ and $h_\b$.  

To verify that the parent Lagrangian \eqref{LagParent} is indeed equivalent to the starting Lagrangian \eqref{Lag1} we perform the variation with respect to dual fields that we have introduced. Varying with respect to $B^\k$ and $H^\k$ we find that
\ba
&
 U{}^{\k}{} 
 + \fr12 Da{}^{a} \we Da{}^{b} \inv{}{}^{\k}{}^{\a}M{}_{\a}{}_{a}{}_{b} 
-  Da{}^{a}{}\we Db{}_{b}{} \inv{}{}^{\k}{}^{\a}M{}_{\a}{}_{a}{}^{b} 
 + \fr12 Db{}_{a}{} \we Db{}_{b}{} \inv{}{}^{\k}{}^{\a}M{}_{\a}{}^{a}{}^{b} 
\nn \\ &
- * D\x{}_{\h}{} G^{\h \l} \Pperp{}^{\k}_{\l}
 - * Da{}^{a}{}H{}_{a}{}^{\l}\Pperp{}^{\k}_{\l} 
 + * Db{}_{a}{} G{}^{a}{}^{\l}\Pperp{}^{\k}_{\l}
- * D\til\x{}^{\h}{} H{}_{\h}{}^{\l}\Pperp{}^{\k}_{\l}
\nn \\& 
- \til\x{}^{\h}\Pi{}^{\k}_{\l}F{}^{\ihat}{} M{}_{\ihat}{}_{\h}{}^{\l} 
-  G{}_{\a}{}_{\b}\inv{}{}^{\k}{}^{\a} d* F{}^{\b}
-  G{}_{\a}{}_{\ihat}\inv{}{}^{\k}{}^{\a} d * F{}^{\ihat}
= 0 \, , 
\label{Beom}
\ea
where the $\Pi_\l^\k$ projection of this equation is obtained from the variation with respect to $B^\k$ and the $\Pperp_\l^\k$ projection is obtained from the variation with respect to $H^\k$. Similarly varying with respect to $\r_\b$ and $h_\b$ gives  
\ba
 \hat D \til\r{}_{\a}+  G{}_{\a}{}_{\b}* F{}^{\b}{} + G{}_{\a}{}_{\ihat}* F{}^{\ihat}  = 0 \,  , 
 \label{Rhoeom}
\ea
where the $\Pi_\a^\b$ projection comes from the variation with respect to $\r_\a$ and the $\Pperp_\a^\b$ projection comes from the variation with respect to $h_\b$.

Substituting these equations into \eqref{LagParent} and making use of certain total derivative identities we return to the original Lagrangian \eqref{Lag1}. This identifies that the Lagrangian \eqref{LagParent} represents an appropriate parent Lagrangian with which to perform the dualisation of  \eqref{Lag1}.

Next we may consider varying the action \eqref{LagParent} with respect to the old variables $\x_\k$ and $A^\a$. The variation with respect to $A^\a$ is most easily understood by splitting it into its   $\Pi_\a^\b$ and $\Pperp_\a^\b$ projections.
The $\Pi_\a^\b$ projection contracted with $\inv^{\a \k}$ gives 
\ba
& 
D\x{}_{\h} G{}^{\h}{}^{\l}  \Pi^{\k}_{\l} + D\til\x{}^{\h} H{}_{\h}{}^{\l}  \Pi^{\k}_{\l}
+ Da{}^{a}{} H{}_{a}{}^{\l} \Pi^{\k}_{\l} - Db{}_{a} G{}^{a}{}^{\l}\Pi^{\k}_{\l} 
+ \Pi^{\k}_{\l} * U{}^{\l} = 0 \, , 
 \label{xieom}
\ea
while the $\Pperp_\a^\b$ projection gives a Bianchi identity for $h_\b$ which implies that
\ba
\hat D \rho_\a & = d \r_\b - \fr12 k_{\b \k} B^\k - \fr12 M_{\a a b} a^a D a^b 
 + \fr12 M_{\a a}{}^{ b} a^a D b_b 
 \nn \\&
 -  \fr12 M_{\a a}{}^{ b} b_b D a^a  -  \fr12 M_{\a}{}^{ a b} b_a D b_b \, , 
\label{DrhoAsCovD}
\ea
Similarly the variation of \eqref{LagParent} with respect to $\x_\k$ is most easily understood by considering its projections with respect to $\Pi_\l^\k$  and $\Pperp_\l^\k$. The $\Pi_\l^\k$ projection gives an equation which represents the derivative of \eqref{xieom} so imposes no additional constraint. Alternatively, the $\Pperp_\l^\k$ projection implies a Bianchi identity for $H^\k$ which is solved by 
\ba
U^\k & = d B^\l  +  \fr12 M_{\ihat a}{}^\k F^\ihat a^a  + \fr12 M_{\ihat a}{}^\k  A^\ihat Da^a 
\nn \\& 
+  \fr12 M_{\ihat }{}^{a \k} F^\ihat b_a  + \fr12 M_{\ihat }{}^{a \k}  A^\ihat Db_a + M_{\ihat \l}{}^\k F^\ihat \til \x^\l  \, . 
\label{UAsCovD}
\ea
Finally we may form a further useful equation by taking the exterior derivative of \eqref{xieom} and contracting with $\inv^{\a \k}$, which gives
\ba
& 
 \Pi{}^{\a}_\b F{}^{\b} 
- G^{-1}{}{}_{\k}{}_{\l}H{}_{a}{}^{\k}\inv{}{}^{\l}{}^{\a} d Da{}^{a}
- G^{-1}{}{}_{\l}{}_{\h}H{}_{\k}{}^{\l}\inv{}{}^{\h}{}^{\a} d D\til\x{}^{\k} 
\nn \\ & 
+ G{}^{a}{}^{\l}G^{-1}{}{}_{\k}{}_{\l}\inv{}{}^{\k}{}^{\a}d Db{}_{a}
- G^{-1}{}{}_{\k}{}_{\l}\inv{}{}^{\k}{}^{\a} d * U{}^{\l} = 0 \, . 
 \label{FConstraint}
\ea

Then splitting $D \xi_\k$ and $F^\a$ into their two projections in \eqref{LagParent} we may use \eqref{xieom} to eliminate $\Pi_{\k}^\l D \xi_\l$  and \eqref{FConstraint} to eliminate $\Pi_{\b}^\a F^\b$. We may then use \eqref{DrhoAsCovD}, \eqref{UAsCovD} and various total derivative identities to eliminate the remaining projections $\Pperp_{k}^\l D \xi_\l$ and $\Pperp_{\b}^\a F^\b$. Having done this we arrive at the dual Lagrangian
\ba
S  =\,\,& \fr{1}{2} \int \bls  R  * 1  - \fr12 G_{\ihat \jhat} d L^\ihat \we * d L^\jhat- \fr12 G_{\a \b} d L^\a \we * d L^\b- G_{\ihat \a} d L^\ihat \we * d L^\a \\\nn
& - \fr{1}{2}\til G{}_{a}{}_{b} Da{}^{a}\w*Da{}^{b}- \fr{1}{2}\til G{}^{a}{}^{b} Db{}_{a}\w*Db{}_{b} +\til H{}_{a}{}^{b}Da{}^{a}\w*Db{}_{b}  - \fr{1}{2}G^{-1}{}{}^{\a}{}^{\b}\hat D \r{}_{\a}\w*\hat D \r{}_{\b}  \\\nn
& - \fr{1}{2}\til G{}_{\k}{}_{\l} D\til\x{}^{\k}\w*D\til\x{}^{\l} -  \til G{}_{a}{}_{\k}Da{}^{a}\w*D\til\x{}^{\k} +\til H{}_{\k}{}^{a}Db{}_{a}\w*D\til\x{}^{\k}- G^{-1}{}{}_{\k}{}_{\l}G{}^{a}{}^{\k}Db{}_{a}\w U{}^{\l} \\\nn
&  + G^{-1}{}{}_{\k}{}_{\l}H{}_{a}{}^{\k}Da{}^{a}{}\w U{}^{\l} +G^{-1}{}{}_{\l}{}_{\h}H{}_{\k}{}^{\l} D\til\x{}^{\k}\w U{}^{\h} - G^{-1}{}{}^{\a}{}^{\b}G{}_{\b}{}_{\ihat}\hat D \r{}_{\a}\w F{}^{\ihat}  \\\nn
&- \fr{1}{3}M{}_{\ihat}{}_{a}{}_{\k}A{}^{\ihat}\w Da{}^{a}\w D\til\x{}^{\k} - \fr{1}{3}M{}_{\ihat}{}_{\k}{}^{a}A{}^{\ihat}\w Db{}_{a}\w D\til\x{}^{\k} + \fr{1}{3}M{}_{\ihat}{}_{a}{}^{b}A{}^{\ihat}\w Da{}^{a}\w Db{}_{b}\\\nn
& - \fr{1}{2}G^{-1}{}{}_{\k}{}_{\l}U{}^{\k}\w*U{}^{\l} - \fr{1}{2}G{}_{\ihat}{}_{\jhat}F{}^{\ihat}\w*F{}^{\jhat} - \fr{1}{3}M{}_{\ihat}{}_{a}{}_{b}a{}^{a}Da{}^{b}\w F{}^{\ihat}+ \fr{1}{2}G^{-1}{}{}^{\a}{}^{\b}G{}_{\a}{}_{\ihat}G{}_{\b}{}_{\jhat}F{}^{\ihat}\w*F{}^{\jhat} \\\nn
& + \fr{1}{3}M{}_{\ihat}{}_{a}{}^{b}a{}^{a}Db{}_{b}\w F{}^{\ihat} - \fr{1}{3}M{}_{\ihat}{}_{a}{}_{\k}a{}^{a}D\til\x{}^{\k}\w F{}^{\ihat} - \fr{1}{6}M{}_{\ihat}{}_{a}{}_{b}A{}^{\ihat}\w Da{}^{a}\w Da{}^{b}\\\nn
& - \fr{1}{6}M{}_{\ihat}{}^{a}{}^{b}A{}^{\ihat}\w Db{}_{a}\w Db{}_{b} - \fr{1}{3}M{}_{\ihat}{}_{b}{}^{a}b{}_{a}Da{}^{b}\w F{}^{\ihat} - \fr{1}{3}M{}_{\ihat}{}^{a}{}^{b}b{}_{a}Db{}_{b}\w F{}^{\ihat}\\\nn
& - \fr{1}{3}M{}_{\ihat}{}_{\k}{}^{a}b{}_{a}D\til\x{}^{\k}\w F{}^{\ihat} + \fr{1}{3}M{}_{\ihat}{}_{a}{}_{\k}\til\x{}^{\k}Da{}^{a}\w F{}^{\ihat} + \fr{1}{3}M{}_{\ihat}{}_{\k}{}^{a}\til\x{}^{\k}Db{}_{a}\w F{}^{\ihat}+ V*1 \brs \, .
\label{LagDual} 
\ea
We may then make the symmetries of \eqref{LagDual} more transparent by making the field redefinition 
\ba
B^\k = A^\k + \fr12 M_{\ihat a}{}^\k A^\ihat a^a - \fr12 M_{\ihat}{}^{a \k} A^\ihat b_a \, . 
\ea
This allow $U^\k$ and $\hat D \rho_\a $ to be written as shown in \eqref{hat}.

\bibliographystyle{JHEP}
%\bibliography{refs}

\begin{thebibliography}{99}
%%%%%%%%%%%%%%%%%%%%%%%%%%%%%%%%%%%%%%%%%%%%%%%%
%\cite{Vafa:1996xn}
\bibitem{Vafa:1996xn} 
  C.~Vafa,
  ``Evidence for F theory,''
  Nucl.\ Phys.\ B {\bf 469}, 403 (1996)
  [hep-th/9602022].
  %%CITATION = HEP-TH/9602022;%%
  %1009 citations counted in INSPIRE as of 23 Apr 2015


%\cite{Hayashi:2014kca}
\bibitem{Hayashi:2014kca} 
  H.~Hayashi, C.~Lawrie, D.~R.~Morrison and S.~Sch\"afer-Nameki,
  ``Box Graphs and Singular Fibers,''
  JHEP {\bf 1405}, 048 (2014)
  [arXiv:1402.2653 [hep-th]].
  %%CITATION = ARXIV:1402.2653;%%
  %20 citations counted in INSPIRE as of 23 Apr 2015


%\cite{Braun:2014kla}
\bibitem{Braun:2014kla} 
  A.~P.~Braun and S.~Sch\"afer-Nameki,
  ``Box Graphs and Resolutions I,''
  arXiv:1407.3520 [hep-th].
  %%CITATION = ARXIV:1407.3520;%%
  %9 citations counted in INSPIRE as of 23 Apr 2015


%\cite{Esole:2014bka}
\bibitem{Esole:2014bka} 
  M.~Esole, S.~H.~Shao and S.~T.~Yau,
  ``Singularities and Gauge Theory Phases,''
  arXiv:1402.6331 [hep-th].
  %%CITATION = ARXIV:1402.6331;%%
  %13 citations counted in INSPIRE as of 23 Apr 2015


%\cite{Esole:2014hya}
\bibitem{Esole:2014hya} 
  M.~Esole, S.~H.~Shao and S.~T.~Yau,
  ``Singularities and Gauge Theory Phases II,''
  arXiv:1407.1867 [hep-th].
  %%CITATION = ARXIV:1407.1867;%%
  %8 citations counted in INSPIRE as of 23 Apr 2015


%\cite{Grimm:2010ez}
\bibitem{Grimm:2010ez} 
  T.~W.~Grimm and T.~Weigand,
  ``On Abelian Gauge Symmetries and Proton Decay in Global F-theory GUTs,''
  Phys.\ Rev.\ D {\bf 82}, 086009 (2010)
  [arXiv:1006.0226 [hep-th]].
  %%CITATION = ARXIV:1006.0226;%%
  %113 citations counted in INSPIRE as of 23 Apr 2015


%\cite{Morrison:2012ei}
\bibitem{Morrison:2012ei} 
  D.~R.~Morrison and D.~S.~Park,
  ``F-Theory and the Mordell-Weil Group of Elliptically-Fibered Calabi-Yau Threefolds,''
  JHEP {\bf 1210}, 128 (2012)
  [arXiv:1208.2695 [hep-th]].
  %%CITATION = ARXIV:1208.2695;%%
  %50 citations counted in INSPIRE as of 23 Apr 2015


%\cite{Braun:2013yti}
\bibitem{Braun:2013yti} 
  V.~Braun, T.~W.~Grimm and J.~Keitel,
  ``New Global F-theory GUTs with U(1) symmetries,''
  JHEP {\bf 1309}, 154 (2013)
  [arXiv:1302.1854 [hep-th]].
  %%CITATION = ARXIV:1302.1854;%%
  %41 citations counted in INSPIRE as of 23 Apr 2015


%\cite{Grimm:2013oga}
\bibitem{Grimm:2013oga} 
  T.~W.~Grimm, A.~Kapfer and J.~Keitel,
  ``Effective action of 6D F-Theory with U(1) factors: Rational sections make Chern-Simons terms jump,''
  JHEP {\bf 1307}, 115 (2013)
  [arXiv:1305.1929 [hep-th]].
  %%CITATION = ARXIV:1305.1929;%%
  %24 citations counted in INSPIRE as of 23 Apr 2015


%\cite{Kuntzler:2014ila}
\bibitem{Kuntzler:2014ila} 
  M.~Kuntzler and S.~Sch\"afer-Nameki,
  ``Tate Trees for Elliptic Fibrations with Rank one Mordell-Weil group,''
  arXiv:1406.5174 [hep-th].
  %%CITATION = ARXIV:1406.5174;%%
  %11 citations counted in INSPIRE as of 23 Apr 2015


%\cite{Borchmann:2013hta}
\bibitem{Borchmann:2013hta} 
  J.~Borchmann, C.~Mayrhofer, E.~Palti and T.~Weigand,
  ``SU(5) Tops with Multiple U(1)s in F-theory,''
  Nucl.\ Phys.\ B {\bf 882}, 1 (2014)
  [arXiv:1307.2902 [hep-th]].
  %%CITATION = ARXIV:1307.2902;%%
  %35 citations counted in INSPIRE as of 23 Apr 2015


%\cite{Cvetic:2013nia}
\bibitem{Cvetic:2013nia} 
  M.~Cveti\v{c}, D.~Klevers and H.~Piragua,
  ``F-Theory Compactifications with Multiple U(1)-Factors: Constructing Elliptic Fibrations with Rational Sections,''
  JHEP {\bf 1306}, 067 (2013)
  [arXiv:1303.6970 [hep-th]].
  %%CITATION = ARXIV:1303.6970;%%
  %45 citations counted in INSPIRE as of 23 Apr 2015


%\cite{Cvetic:2013uta}
\bibitem{Cvetic:2013uta} 
  M.~Cveti\v{c}, A.~Grassi, D.~Klevers and H.~Piragua,
  ``Chiral Four-Dimensional F-Theory Compactifications With SU(5) and Multiple U(1)-Factors,''
  JHEP {\bf 1404}, 010 (2014)
  [arXiv:1306.3987 [hep-th]].
  %%CITATION = ARXIV:1306.3987;%%
  %37 citations counted in INSPIRE as of 23 Apr 2015


%\cite{Borchmann:2013jwa}
\bibitem{Borchmann:2013jwa} 
  J.~Borchmann, C.~Mayrhofer, E.~Palti and T.~Weigand,
  ``Elliptic fibrations for $SU(5)\times U(1)\times U(1)$ F-theory vacua,''
  Phys.\ Rev.\ D {\bf 88}, no. 4, 046005 (2013)
  [arXiv:1303.5054 [hep-th]].
  %%CITATION = ARXIV:1303.5054;%%
  %42 citations counted in INSPIRE as of 23 Apr 2015


%\cite{Cvetic:2013qsa}
\bibitem{Cvetic:2013qsa} 
  M.~Cveti\v{c}, D.~Klevers, H.~Piragua and P.~Song,
  ``Elliptic fibrations with rank three Mordell-Weil group: F-theory with U(1) x U(1) x U(1) gauge symmetry,''
  JHEP {\bf 1403}, 021 (2014)
  [arXiv:1310.0463 [hep-th], arXiv:1310.0463].
  %%CITATION = ARXIV:1310.0463;%%
  %33 citations counted in INSPIRE as of 23 Apr 2015


%\cite{Braun:2013nqa}
\bibitem{Braun:2013nqa} 
  V.~Braun, T.~W.~Grimm and J.~Keitel,
  ``Geometric Engineering in Toric F-Theory and GUTs with U(1) Gauge Factors,''
  JHEP {\bf 1312}, 069 (2013)
  [arXiv:1306.0577 [hep-th]].
  %%CITATION = ARXIV:1306.0577;%%
  %38 citations counted in INSPIRE as of 23 Apr 2015


%\cite{Klevers:2014bqa}
\bibitem{Klevers:2014bqa} 
  D.~Klevers, D.~K.~Mayorga Pe\~na, P.~K.~Oehlmann, H.~Piragua and J.~Reuter,
  ``F-Theory on all Toric Hypersurface Fibrations and its Higgs Branches,''
  JHEP {\bf 1501}, 142 (2015)
  [arXiv:1408.4808 [hep-th]].
  %%CITATION = ARXIV:1408.4808;%%
  %18 citations counted in INSPIRE as of 23 Apr 2015


%\cite{Braun:2014qka}
\bibitem{Braun:2014qka} 
  V.~Braun, T.~W.~Grimm and J.~Keitel,
  ``Complete Intersection Fibers in F-Theory,''
  JHEP {\bf 1503}, 125 (2015)
  [arXiv:1411.2615 [hep-th]].
  %%CITATION = ARXIV:1411.2615;%%
  %8 citations counted in INSPIRE as of 23 Apr 2015


%\cite{Lawrie:2014uya}
\bibitem{Lawrie:2014uya} 
  C.~Lawrie and D.~Sacco,
  ``Tate's algorithm for F-theory GUTs with two U(1)s,''
  JHEP {\bf 1503}, 055 (2015)
  [arXiv:1412.4125 [hep-th]].
  %%CITATION = ARXIV:1412.4125;%%
  %5 citations counted in INSPIRE as of 23 Apr 2015


%\cite{Lawrie:2015hia}
\bibitem{Lawrie:2015hia} 
  C.~Lawrie, S.~Sch\"afer-Nameki and J.~M.~Wong,
  ``F-theory and All Things Rational: Surveying U(1) Symmetries with Rational Sections,''
  arXiv:1504.05593 [hep-th].
  %%CITATION = ARXIV:1504.05593;%%


%\cite{Grimm:2015zea}
\bibitem{Grimm:2015zea} 
  T.~W.~Grimm and A.~Kapfer,
  ``Anomaly Cancelation in Field Theory and F-theory on a Circle,''
  arXiv:1502.05398 [hep-th].
  %%CITATION = ARXIV:1502.05398;%%
  %4 citations counted in INSPIRE as of 23 Apr 2015


%\cite{Braun:2014oya}
\bibitem{Braun:2014oya} 
  V.~Braun and D.~R.~Morrison,
  ``F-theory on Genus-One Fibrations,''
  JHEP {\bf 1408}, 132 (2014)
  [arXiv:1401.7844 [hep-th]].
  %%CITATION = ARXIV:1401.7844;%%
  %22 citations counted in INSPIRE as of 23 Apr 2015


%\cite{Morrison:2014era}
\bibitem{Morrison:2014era} 
  D.~R.~Morrison and W.~Taylor,
  ``Sections, multisections, and U(1) fields in F-theory,''
  arXiv:1404.1527 [hep-th].
  %%CITATION = ARXIV:1404.1527;%%
  %21 citations counted in INSPIRE as of 23 Apr 2015


%\cite{Anderson:2014yva}
\bibitem{Anderson:2014yva} 
  L.~B.~Anderson, I.~Garc\'ia-Etxebarria, T.~W.~Grimm and J.~Keitel,
  ``Physics of F-theory compactifications without section,''
  JHEP {\bf 1412}, 156 (2014)
  [arXiv:1406.5180 [hep-th]].
  %%CITATION = ARXIV:1406.5180;%%
  %13 citations counted in INSPIRE as of 23 Apr 2015


%\cite{Garcia-Etxebarria:2014qua}
\bibitem{Garcia-Etxebarria:2014qua} 
  I.~Garc\'ia-Etxebarria, T.~W.~Grimm and J.~Keitel,
  ``Yukawas and discrete symmetries in F-theory compactifications without section,''
  JHEP {\bf 1411}, 125 (2014)
  [arXiv:1408.6448 [hep-th]].
  %%CITATION = ARXIV:1408.6448;%%
  %15 citations counted in INSPIRE as of 23 Apr 2015


%\cite{Mayrhofer:2014haa}
\bibitem{Mayrhofer:2014haa} 
  C.~Mayrhofer, E.~Palti, O.~Till and T.~Weigand,
  ``Discrete Gauge Symmetries by Higgsing in four-dimensional F-Theory Compactifications,''
  JHEP {\bf 1412}, 068 (2014)
  [arXiv:1408.6831 [hep-th]].
  %%CITATION = ARXIV:1408.6831;%%
  %13 citations counted in INSPIRE as of 23 Apr 2015


%\cite{Mayrhofer:2014laa}
\bibitem{Mayrhofer:2014laa} 
  C.~Mayrhofer, E.~Palti, O.~Till and T.~Weigand,
  ``On Discrete Symmetries and Torsion Homology in F-Theory,''
  arXiv:1410.7814 [hep-th].
  %%CITATION = ARXIV:1410.7814;%%
  %10 citations counted in INSPIRE as of 23 Apr 2015


%\cite{Cvetic:2015moa}
\bibitem{Cvetic:2015moa} 
  M.~Cveti\v{c}, R.~Donagi, D.~Klevers, H.~Piragua and M.~Poretschkin,
  ``F-Theory Vacua with $Z_3$ Gauge Symmetry,''
  arXiv:1502.06953 [hep-th].
  %%CITATION = ARXIV:1502.06953;%%
  %3 citations counted in INSPIRE as of 23 Apr 2015


%\cite{Douglas:2014ywa}
\bibitem{Douglas:2014ywa} 
  M.~R.~Douglas, D.~S.~Park and C.~Schnell,
  ``The Cremmer-Scherk Mechanism in F-theory Compactifications on K3 Manifolds,''
  JHEP {\bf 1405}, 135 (2014)
  [arXiv:1403.1595 [hep-th]].
  %%CITATION = ARXIV:1403.1595;%%
  %8 citations counted in INSPIRE as of 23 Apr 2015


%\cite{Banks:2010zn}
\bibitem{Banks:2010zn} 
  T.~Banks and N.~Seiberg,
  ``Symmetries and Strings in Field Theory and Gravity,''
  Phys.\ Rev.\ D {\bf 83}, 084019 (2011)
  [arXiv:1011.5120 [hep-th]].
  %%CITATION = ARXIV:1011.5120;%%
  %114 citations counted in INSPIRE as of 23 Apr 2015


%\cite{Grimm:2011tb}
\bibitem{Grimm:2011tb} 
  T.~W.~Grimm, M.~Kerstan, E.~Palti and T.~Weigand,
  ``Massive Abelian Gauge Symmetries and Fluxes in F-theory,''
  JHEP {\bf 1112}, 004 (2011)
  [arXiv:1107.3842 [hep-th]].
  %%CITATION = ARXIV:1107.3842;%%
  %54 citations counted in INSPIRE as of 23 Apr 2015


%\cite{Grimm:2013fua}
\bibitem{Grimm:2013fua} 
  T.~W.~Grimm and T.~G.~Pugh,
  ``Gauged supergravities and their symmetry-breaking vacua in F-theory,''
  JHEP {\bf 1306}, 012 (2013)
  [arXiv:1302.3223 [hep-th]].
  %%CITATION = ARXIV:1302.3223;%%
  %4 citations counted in INSPIRE as of 23 Apr 2015


%\cite{Braun:2014nva}
\bibitem{Braun:2014nva} 
  A.~P.~Braun, A.~Collinucci and R.~Valandro,
  ``The fate of U(1)'s at strong coupling in F-theory,''
  JHEP {\bf 1407}, 028 (2014)
  [arXiv:1402.4054 [hep-th]].
  %%CITATION = ARXIV:1402.4054;%%
  %16 citations counted in INSPIRE as of 23 Apr 2015


%\cite{Jockers:2004yj}
\bibitem{Jockers:2004yj} 
  H.~Jockers and J.~Louis,
  ``The Effective action of D7-branes in N = 1 Calabi-Yau orientifolds,''
  Nucl.\ Phys.\ B {\bf 705}, 167 (2005)
  [hep-th/0409098].
  %%CITATION = HEP-TH/0409098;%%
  %155 citations counted in INSPIRE as of 23 Apr 2015


%\cite{Camara:2011jg}
\bibitem{Camara:2011jg} 
  P.~G.~C\'amara, L.~E.~Ib\'a\~nez and F.~Marchesano,
  ``RR photons,''
  JHEP {\bf 1109}, 110 (2011)
  [arXiv:1106.0060 [hep-th]].
  %%CITATION = ARXIV:1106.0060;%%
  %35 citations counted in INSPIRE as of 23 Apr 2015


%\cite{BerasaluceGonzalez:2012vb}
\bibitem{BerasaluceGonzalez:2012vb} 
  M.~Berasaluce-Gonz\'alez, P.~G.~C\'amara, F.~Marchesano, D.~Regalado and A.~M.~Uranga,
  ``Non-Abelian discrete gauge symmetries in 4d string models,''
  JHEP {\bf 1209}, 059 (2012)
  [arXiv:1206.2383 [hep-th]].
  %%CITATION = ARXIV:1206.2383;%%
  %41 citations counted in INSPIRE as of 23 Apr 2015


%\cite{Cvetic:2015txa}
\bibitem{Cvetic:2015txa} 
  M.~Cveti\v{c}, D.~Klevers, D.~K.~M.~Pe\~na, P.~K.~Oehlmann and J.~Reuter,
  ``Three-Family Particle Physics Models from Global F-theory Compactifications,''
  arXiv:1503.02068 [hep-th].
  %%CITATION = ARXIV:1503.02068;%%


%\cite{Gukov:1998kn}
\bibitem{Gukov:1998kn} 
  S.~Gukov, M.~Rangamani and E.~Witten,
  ``Dibaryons, strings and branes in AdS orbifold models,''
  JHEP {\bf 9812}, 025 (1998)
  [hep-th/9811048].
  %%CITATION = HEP-TH/9811048;%%
  %73 citations counted in INSPIRE as of 23 Apr 2015


%\cite{BerasaluceGonzalez:2011wy}
\bibitem{BerasaluceGonzalez:2011wy} 
  M.~Berasaluce-Gonz\'alez, L.~E.~Ib\'a\~nez, P.~Soler and A.~M.~Uranga,
  ``Discrete gauge symmetries in D-brane models,''
  JHEP {\bf 1112}, 113 (2011)
  [arXiv:1106.4169 [hep-th]].
  %%CITATION = ARXIV:1106.4169;%%
  %40 citations counted in INSPIRE as of 23 Apr 2015


%\cite{Hull:1985pq}
\bibitem{Hull:1985pq} 
  C.~M.~Hull, A.~Karlhede, U.~Lindstr\"om and M.~Ro\v{c}ek,
  ``Nonlinear $\sigma$ Models and Their Gauging in and Out of Superspace,''
  Nucl.\ Phys.\ B {\bf 266}, 1 (1986).
  %%CITATION = NUPHA,B266,1;%%
  %130 citations counted in INSPIRE as of 23 Apr 2015


%\cite{Harvey:2005it}
\bibitem{Harvey:2005it} 
  J.~A.~Harvey,
  ``TASI 2003 lectures on anomalies,''
  hep-th/0509097.
  %%CITATION = HEP-TH/0509097;%%
  %38 citations counted in INSPIRE as of 23 Apr 2015


%\cite{Bilal:2008qx}
\bibitem{Bilal:2008qx} 
  A.~Bilal,
  ``Lectures on Anomalies,''
  arXiv:0802.0634 [hep-th].
  %%CITATION = ARXIV:0802.0634;%%
  %29 citations counted in INSPIRE as of 23 Apr 2015


%\cite{Grimm:2004uq}
\bibitem{Grimm:2004uq} 
  T.~W.~Grimm and J.~Louis,
  ``The Effective action of N = 1 Calabi-Yau orientifolds,''
  Nucl.\ Phys.\ B {\bf 699}, 387 (2004)
  [hep-th/0403067].
  %%CITATION = HEP-TH/0403067;%%
  %229 citations counted in INSPIRE as of 23 Apr 2015


%\cite{Grimm:2010ks}
\bibitem{Grimm:2010ks} 
  T.~W.~Grimm,
  ``The N=1 effective action of F-theory compactifications,''
  Nucl.\ Phys.\ B {\bf 845}, 48 (2011)
  [arXiv:1008.4133 [hep-th]].
  %%CITATION = ARXIV:1008.4133;%%
  %68 citations counted in INSPIRE as of 23 Apr 2015


%\cite{BerasaluceGonzalez:2012zn}
\bibitem{BerasaluceGonzalez:2012zn} 
  M.~Berasaluce-Gonz\'alez, P.~G.~C\'amara, F.~Marchesano and A.~M.~Uranga,
  ``$Z_p$ charged branes in flux compactifications,''
  JHEP {\bf 1304}, 138 (2013)
  [arXiv:1211.5317 [hep-th]].
  %%CITATION = ARXIV:1211.5317;%%
  %15 citations counted in INSPIRE as of 23 Apr 2015


%\cite{Samtleben:2008pe}
\bibitem{Samtleben:2008pe} 
  H.~Samtleben,
  ``Lectures on Gauged Supergravity and Flux Compactifications,''
  Class.\ Quant.\ Grav.\  {\bf 25}, 214002 (2008)
  [arXiv:0808.4076 [hep-th]].
  %%CITATION = ARXIV:0808.4076;%%
  %91 citations counted in INSPIRE as of 23 Apr 2015


%\cite{deWit:2004yr}
\bibitem{deWit:2004yr} 
  B.~de Wit, H.~Nicolai and H.~Samtleben,
  ``Gauged supergravities in three-dimensions: A Panoramic overview,''
  PoS jhw {\bf 2003}, 016 (2003)
  [PoS jhw {\bf 2003}, 018 (2003)]
  [hep-th/0403014].
  %%CITATION = HEP-TH/0403014;%%
  %44 citations counted in INSPIRE as of 23 Apr 2015


%\cite{Abel:2008ai}
\bibitem{Abel:2008ai} 
  S.~A.~Abel, M.~D.~Goodsell, J.~Jaeckel, V.~V.~Khoze and A.~Ringwald,
  ``Kinetic Mixing of the Photon with Hidden U(1)s in String Phenomenology,''
  JHEP {\bf 0807}, 124 (2008)
  [arXiv:0803.1449 [hep-ph]].
  %%CITATION = ARXIV:0803.1449;%%
  %187 citations counted in INSPIRE as of 23 Apr 2015


%\cite{Marchesano:2014bia}
\bibitem{Marchesano:2014bia} 
  F.~Marchesano, D.~Regalado and G.~Zoccarato,
  ``U(1) mixing and D-brane linear equivalence,''
  JHEP {\bf 1408}, 157 (2014)
  [arXiv:1406.2729 [hep-th]].
  %%CITATION = ARXIV:1406.2729;%%
  %2 citations counted in INSPIRE as of 23 Apr 2015


\bibitem{toAppearKinMixing}
T.~W.~Grimm, D.~Regalado and T.~G.~Pugh,
  To appear. 


%\cite{Grimm:2011sk}
\bibitem{Grimm:2011sk} 
  T.~W.~Grimm and R.~Savelli,
  ``Gravitational Instantons and Fluxes from M/F-theory on Calabi-Yau fourfolds,''
  Phys.\ Rev.\ D {\bf 85}, 026003 (2012)
  [arXiv:1109.3191 [hep-th]].
  %%CITATION = ARXIV:1109.3191;%%
  %27 citations counted in INSPIRE as of 23 Apr 2015


%\cite{Grimm:2011fx}
\bibitem{Grimm:2011fx} 
  T.~W.~Grimm and H.~Hayashi,
  ``F-theory fluxes, Chirality and Chern-Simons theories,''
  JHEP {\bf 1203}, 027 (2012)
  [arXiv:1111.1232 [hep-th]].
  %%CITATION = ARXIV:1111.1232;%%
  %71 citations counted in INSPIRE as of 23 Apr 2015


\bibitem{toAppearDuality}
  T.~W.~Grimm, D.~Regalado and T.~G.~Pugh,
  To appear. 

%\cite{Marchesano:2013ega}
\bibitem{Marchesano:2013ega} 
  F.~Marchesano, D.~Regalado and L.~V\'azquez-Mercado,
  ``Discrete flavor symmetries in D-brane models,''
  JHEP {\bf 1309}, 028 (2013)
  [arXiv:1306.1284 [hep-th]].
  %%CITATION = ARXIV:1306.1284;%%
  %20 citations counted in INSPIRE as of 23 Apr 2015
  
  
%%%%%%%%%%%%%%%%%%%%%%%%%%%%%%%%%%%%%%%%%%%%%%%%
\end{thebibliography}
%%%%%%%%%%%%%%%%%%%%%%%%%%%%%%%%%%%%%%%%%%%%%%%%

%%%%%%%%%%%%%%%%%%%%%%%%%%%%%%%%%%%%%%%%%%%%%%%%

\end{document}